\documentclass[AMA,STIX1COL]{WileyNJD-v2}

\articletype{Article Type}%

\received{Day Month 20xx}
\revised{Day Month 20xx}
\accepted{Day Month 20xx}

\usepackage{amsmath, amsfonts, amsthm}
\usepackage{bm, bbm, xcolor, colonequals}
\newcommand{\bfbeta}{\bm{\beta}}
\newcommand{\bfgamma}{\bm{\gamma}}
\newcommand{\bfxi}{\bm{\xi}}
\newcommand{\bfU}{\bm{U}}
\newcommand{\bfW}{\bm{W}}

\begin{document}

\title{Estimation of Treatment Effect in Clinical Trials of Continuous Endpoints with Retrieved Dropouts}

\author[1]{Myeongjong Kang}

\author[2]{Sangyoon Yi}

\authormark{Kang and Yi}

\address[1]{\orgname{Merck \& Co., Inc.}, \orgaddress{\state{Rahway, New Jersey}, \country{USA}}}

\address[2]{\orgdiv{Department of Statistics}, \orgname{Oklahoma State University}, \orgaddress{Stillwater, Oklahoma, \country{USA}}}
 
\corres{*Sangyoon Yi, \email{sayi@okstate.edu}}

\abstract[Summary]{The estimand framework provides guidance on handling intercurrent events, such as treatment discontinuation, in the analysis of clinical trial responses. Under ICH E9(R1), the treatment policy (TP) strategy incorporates post-discontinuation data to reflect treatment effects in real-world practice. However, many existing approaches focus primarily on imputing missing endpoint values for lost-to-follow-up subjects and do not explicitly model completers, retrieved dropouts (RDs), and lost-to-follow-up subjects within a unified framework. This may obscure the relationship between modeling assumptions and the estimand of interest when RD data are present. We propose a likelihood-based model for continuous endpoints that integrates data from all subject categories, including RDs. The approach combines an analysis of covariance formulation with a probit model for treatment discontinuation, enabling explicit formulation of treatment effects for estimands defined using the hypothetical and TP strategies. Estimation is carried out via a computationally efficient maximum likelihood procedure. Numerical studies demonstrate that the proposed method achieves improved bias and variability properties compared with commonly used imputation-based approaches.}


\maketitle

\section{Introduction}

The choice of a missing data handling approach in clinical trials is closely linked to how the estimand defines the impact of intercurrent events (IEs) \citep{ICH2019E9R1}. As outlined in the ICH E9(R1) addendum, an estimand specifies the treatment effect of interest by explicitly addressing IEs such as treatment discontinuation or use of rescue medication, which may lead to missing or post-IE data. These IEs should be considered when selecting an appropriate missing data method to ensure alignment with the estimand's objectives and to obtain an accurate estimate of the treatment effect. Many conventional missing data methods \citep{Molenberghs2007, Little2019} used in clinical trials require adaptation to comply with the ICH E9(R1) addendum, particularly when the treatment policy (TP) strategy is adopted, as such methods are often more naturally aligned with a hypothetical strategy \citep{Fletcher2022}. In response, methods aligned with the TP strategy have been proposed \citep{Wang2023} and refined, including return-to-baseline (RTB) imputation, washout imputation, and reference-based approaches \citep{Carpenter2013}. However, there remains limited statistical research on unbiased estimation when the TP strategy is employed \citep{Fletcher2022}, highlighting the need for further methodological advancement.

Incorporating data from retrieved dropouts (RDs)--subjects who discontinue treatment but return for endpoint assessment--is essential for accurately characterizing post-discontinuation responses under the TP strategy. It has been shown that the assumption underlying the last observation on-treatment carried forward method may be inappropriate in weight management trials, since subjects in the experimental arm often regained weight after stopping treatment \cite{McEvoy2016}. To address such limitations, regression-based RD multiple imputation approaches have been proposed \citep{Wharton2021, Wang2022}. These methods use observed data from comparable RDs to impute missing endpoint values for subjects who are lost to follow-up or discontinue the study before endpoint assessment (hereafter referred to as LTFUs) and have been shown in simulation studies to reduce bias relative to alternative approaches \citep{Wang2023}. However, their performance depends on the availability and representativeness of RD data, and variance may increase when RD observations are scarce \citep{Wang2023, Bell2025}. More recent work has emphasized the importance of aligning modeling assumptions with estimand attributes in the presence of RD data, including estimation approaches under alternative IE handling strategies \citep{Bell2025, Drury2025}, as well as multiple-imputation approaches that explicitly distinguish on-treatment and off-treatment outcomes \cite{drury2024estimation}. While promising, most existing RD-based approaches primarily rely on imputation of missing endpoint values and do not explicitly model completers, LTFUs, and RDs within a unified likelihood framework. In particular, the formal connection between estimators aligned with the TP strategy and those targeting the hypothetical strategy remains less explicitly developed. The approach proposed in this paper addresses this gap by jointly modeling the endpoint and discontinuation processes within a likelihood-based formulation, thereby clarifying the relationship between modeling assumptions and estimand attributes.

We propose a novel model for continuous endpoints that integrates data from completers, LTFUs, and RDs, and develops a computationally efficient maximum likelihood (ML)-based estimation procedure for key parameters, including the treatment effect under both the hypothetical and TP strategies. The method combines an analysis of covariance (ANCOVA) model for endpoint analysis with a probit model for treatment discontinuation, enabling a plug-in estimator for the TP-strategy-aligned treatment effect. Simulation studies demonstrate that the proposed approach outperforms existing imputation-based methods under the TP strategy, particularly surpassing RD imputation when sufficient RD data are not available. To our knowledge, this is one of the first approaches to explicitly formulate a TP-strategy-aligned treatment effect while clarifying its relationship to the hypothetical-strategy-aligned treatment effect. By transparently addressing IEs, the method ensures alignment with the chosen IE strategy. Using causal inference principles \citep{Lipkovich2020}, it accounts for treatment discontinuation by modeling the discontinuation process as a function of baseline covariates--for example, allowing for higher discontinuation risk among subjects with more severe conditions at baseline, potentially depending on treatment assignment. The model is flexible and extensible; for instance, the probit model can be replaced with a logistic model.

The remainder of this article is organized as follows. Section~\ref{sec:method} introduces our joint model that integrates data from completers, LTFUs, and RDs. Section~\ref{subsec:setup} defines the notations used for model formulation, followed by Section~\ref{subsec:model}, which presents the joint model itself. Section~\ref{subsec:causal} derives the treatment effects under the hypothetical and TP strategies induced by the model and discusses their mathematical connections. Section~\ref{subsec:infer} describes the ML-based inference procedure and outlines the key parameters, including treatment effects under both IE strategies. In Section~\ref{sec:simul}, we assess the performance of the proposed method through simulation studies, comparing it to existing imputation approaches. In Section~\ref{sec:realdata}, we illustrate the proposed method using an application inspired by a publicly available antidepressant dataset. Finally, Section~\ref{sec:concl} concludes the paper and discusses directions for future research. Additional model formulation, likelihood derivations, implementation details for competing imputation methods, and additional simulation results are provided in Supplementary Material. The code for implementing our method and reproducing the figures is available at \url{https://github.com/myeongjong/RDancova}.

\section{Methodology}\label{sec:method} 

\subsection{Setup}\label{subsec:setup}

A total of $N$ subjects are randomized to two treatment groups: placebo and experimental arm. Let $\{Y_{i,t}\}_{t=0}^{T}$ denote the collection of clinical responses of interest (e.g., weight in a weight management trial) measured over $T+1$ visits for the $i$th subject. 
Although repeated measurements are collected, we focus on the baseline ($t=0$) and final visit ($t=T$), consistent with a standard ANCOVA \citep{Montgomery2017}. We define the endpoint as the change from baseline,
\[
Z_{i,T} = Y_{i,T} - Y_{i,0}.
\]
While we present the method using this change-from-baseline endpoint for clarity, the proposed approach can be adapted to other types of endpoints, see \ref{subsec:unif}. Here, $Z_{i,T}$ denotes the endpoint that would be defined at $T$ for each randomized subject; whether this endpoint is observed is determined separately by the indicator $Q_i$ introduced below. The variables $Z_{i,T}$ are assumed to be independent across subjects, reflecting the randomized controlled trial setting. For simplicity, we assume that there are no missing values at baseline and that any missing values occur only at $T$. This assumption is reasonable, since the proportion of subjects with missing baseline data is typically zero or negligible in modern clinical trials, where screening and randomization are required prior to treatment initiation.

We consider treatment discontinuation as the IE of primary interest in this paper. We classify subjects into three categories. Completers remain on treatment and in the study through the endpoint assessment at $T$. LTFU subjects discontinue both treatment and study participation prematurely and therefore have no endpoint assessment at $T$. RD subjects discontinue treatment but remain in the study and return for the endpoint assessment at $T$. Classical missing-data frameworks typically distinguish only between completers and LTFUs. In contrast, our setup explicitly recognizes the presence of RDs, who provide post-treatment-discontinuation responses that may differ from their potential on-treatment responses. Under the TP strategy, post-discontinuation data are included in the endpoint analysis. However, combining RD and completer data without explicitly accounting for their different data-generating mechanisms (e.g., by applying a conventional ANCOVA model with a classical imputation method to the combined data) may oversimplify the structure of the data. In particular, such an approach obscures potential differences between post-discontinuation responses and on-treatment responses at the final visit.

The study or treatment completion status of these three categories can be formalized using two binary indicators. Let $D_{i}$ denote the treatment continuation indicator for $i$th subject:
\[
D_{i} =
\begin{cases}
1, & \text{if the $i$th subject remains on treatment until endpoint assessment at $T$}, \\
0, & \text{otherwise}.
\end{cases}
\]
Among subjects with $D_{i}=0$, let $Q_i$ denote the retrieval indicator for the final-visit response:
\[
Q_i =
\begin{cases}
1, & \text{if the final-visit response is retrieved for endpoint assessment}, \\
0, & \text{if the final-visit response is not observed}.
\end{cases}
\]
Accordingly, subjects are categorized as illustrated in Figure \ref{fig:casevars}. We assume that the case $D_{i}=1$ and $Q_{i}=0$ is implausible, since subjects who continue treatment typically remain in the study for medication dispensing and safety monitoring.

\begin{figure}[t]
        \centering
        \vspace{1mm}
        \def\mca#1{\multicolumn{1}{c}{#1}}
        \begin{tabular}{>{\centering\arraybackslash}p{0.1\textwidth}|>{\centering}p{0.2\textwidth}|>{\centering\arraybackslash}p{0.2\textwidth}|}
        \mca{}  & \mca{$D_i=1$} & \mca{$D_i=0$} \tabularnewline[10pt]\cline{2-3}
        \begin{center}$Q_i= 1$\end{center}      & 
        \begin{center}\text{Completer} \end{center}      & 
        \begin{center}\text{RD}\end{center}    \tabularnewline[30pt]\cline{2-3}
        \begin{center}$Q_i= 0$\end{center}      & 
        \begin{center}Implausible\end{center}   & 
        \begin{center}\text{LTFU}\end{center}    \tabularnewline[30pt]\cline{2-3}
    \end{tabular}
    \caption{Schematic illustration of the relationship between variables $D_i$ and $Q_i$} 
    \label{fig:casevars}
\end{figure}

Unlike classical missing-data settings, RDs contribute observed data that arise after treatment discontinuation. The primary issue is therefore not solely missingness, but the potential change in response distribution following discontinuation. Treating post-discontinuation outcomes as if they were on-treatment outcomes may not align with the selected strategy. This motivates the joint modeling of endpoint and discontinuation processes adopted in our framework.

\subsection{Model}\label{subsec:model}

We model the endpoint $Z_{i,T}$ conditional on baseline measurement $Y_{i,0}$ and covariates through an ANCOVA formulation \citep{Montgomery2017}:
\begin{equation}\label{eqn:resmod}
    Z_{i,T} = \mathbf{W}_{i}^\top \boldsymbol{\beta} + \mathbf{1}\{D_i=0\}\cdot \delta + \epsilon_{i},
\end{equation}
where $\mathbf{W}_{i}^\top = [1\; Y_{i,0}\; \mathbf{X}_{i}^\top]$, $\mathbf{X}_i$ denotes the $i$th subject's fully observed covariate vector including treatment assignment (0 for placebo and 1 for experimental arm), $\boldsymbol{\beta}^\top=[\beta_{0}\ \beta_{\text{base}}\ \mathbf{\beta}_{\mathbf{X}}^\top]$ is the corresponding regression coefficient vector, and $\epsilon_{i} \stackrel{i.i.d.}{\sim} N(0,\sigma_{\epsilon}^{2})$ is the random error. In this formulation, $Z_{i,T}$ represents the endpoint value at $T$ for $i$th subject, whereas its observed value is available only when $Q_i=1$. Thus, our model \eqref{eqn:resmod} characterizes the endpoint distribution associated with treatment discontinuation through $\delta$, while observation of that endpoint is handled separately through the retrieval mechanism.

The parameter $\delta$ represents a mean shift in the endpoint associated with treatment discontinuation before $T$, capturing potential differences between on-treatment and post-discontinuation responses. Correct specification of this post-discontinuation component is therefore important for interpretation of treatment effects under the proposed model. Accordingly, for subjects with $D_i=0$ and $Q_i=0$, model \eqref{eqn:resmod} pertains to the endpoint that would arise after discontinuation, even though the endpoint is not observed.

The treatment continuation indicator $D_i$ and retrieval indicator $Q_i$ are modeled through a nested structure. We specify a probit model for $D_i$:
\begin{equation}\label{mod:probit}
    D_{i} = \mathbf{1} \left\{ \mathbf{W}_{i}^\top \boldsymbol{\gamma} + \eta_{i} < 0 \right\},
\end{equation}
where $\boldsymbol{\gamma}^\top=[\gamma_{0}\ 1\ \boldsymbol{\gamma}_{\mathbf{X}}^\top]$ is a coefficient vector and $\eta_{i} \stackrel{\mathrm{i.i.d.}}{\sim} N(0,1)$ is independent of $\epsilon_i$. Here, the baseline measurement included in $\mathbf{W}_{i}$ may be standardized, if desired, without changing the formulation of the model. Under this formulation, the probability of treatment discontinuation depends on observed baseline measurements and covariates, allowing discontinuation risk to vary across treatment arms and subject characteristics \citep{ICH2016}. Among subjects with $D_i=0$, we model the retrieval indicator $Q_i$ for the final-visit response as
\begin{equation}\label{eqn:retrieval}
    Q_i \mid (D_i=0) \sim \text{Bernoulli}(\pi),
\end{equation}
where $\pi\in(0,1)$ denotes the probability that the final-visit response is retrieved among subjects who discontinue treatment. For subjects with $D_i=1$, we assume $\Pr(Q_i=1\mid D_i=1)=1$.

Within the model specified above, treatment discontinuation depends on observed baseline covariates and treatment assignment. This reflects common clinical settings in which discontinuation risk may vary across treatment arms--for example, lack of efficacy may be more frequent in the placebo arm, whereas adverse events may be more common in the experimental arm--and may also depend on baseline disease severity. Importantly, treatment discontinuation constitutes an IE that may alter the response distribution (e.g., through the shift parameter $\delta$), whereas classical missing data mechanisms describe the process governing whether a response is observed.

Extensions that allow longitudinal settings, in which discontinuation may depend on intermediate data, and treatment-specific post-discontinuation shifts (via a treatment-by-discontinuation interaction) are outlined in Web Appendices B and C of the Supplementary Material.

\subsection{Treatment Effects under Hypothetical and TP Strategies}\label{subsec:causal} 

The distinction between treatment discontinuation and study discontinuation (i.e., no endpoint assessment), described in Section \ref{subsec:model}, is also useful for clarifying the properties of the treatment effect estimator aligned with the TP strategy and its connection to the estimator for the hypothetical strategy under the potential-outcome framework \cite{Lipkovich2024}. For simplicity, we suppress any additional covariates in this subsection and assume that baseline measurement $Y_{i,0}$ is the only covariate, although the arguments extend directly to settings with additional covariates. Accordingly, we omit the bold notation for $\mathbf{X}$ and $\boldsymbol{\beta}$. Also, we omit the explicit potential outcome expression throughout the paper if needed.

Let $D_{i}(x) \in \{0,1\}$ denote the potential treatment-continuation status at $T$ under treatment assignment $X_i = x$ ($1$ for experimental and $0$ for placebo), where $D_i(x)=1$ indicates that the $i$th subject would remain on treatment through $T$ and $D_i(x)=0$ indicates treatment discontinuation before $T$. Potential outcomes for $Q_i$ do not need to be introduced, since LTFU affects only the observability of the endpoint, rather than the definition of the treatment effect itself. Let $Z_{i,T}(x,d)$ denote the potential endpoint value at $T$ if the $i$th subject were assigned treatment $x\in\{0,1\}$ and had treatment-continuation status $d\in\{0,1\}$ by time $T$. We assume that the potential outcome $Z_{i,T}(x,d)$ exists and is well defined for all combinations $(x, d) \in \{0,1\}^2$.

We additionally adopt the standard causal identification assumptions of consistency, positivity, and (conditional) exchangeability. Specifically, we assume:
\begin{enumerate}
    \item[(A1)] \textit{Consistency implied by the Stable Unit Treatment Value Assumption (SUTVA)}: The observed endpoint equals the potential outcome corresponding to the realized treatment assignment and treatment-continuation status: 
    \begin{equation*}
        Z_{i,T} = Z_{i,T}(X_i, D_i(X_i)).
    \end{equation*}
    \item[(A2)] \textit{Positivity of treatment assignment}: Each treatment arm has positive probability for all possible baseline values:
    \begin{equation*}
        \Pr(X_i = x \mid Y_{i,0} = y) > 0 \quad \text{for} \quad x \in \{0,1\}.
    \end{equation*}
    \item[(A3)] \textit{Positivity for treatment discontinuation}: Treatment discontinuation (or continuation) occurs with positive probability within each treatment arm and baseline measurement:
    \begin{equation*}
        \Pr(D_i = d \mid X_i = x, Y_{i,0} = y) > 0 \quad \text{for} \quad (x,d) \in \{0,1\}^2.
    \end{equation*}
    \item[(A4)] \textit{Exchangeability of treatment assignment}: Conditional on the baseline measurement, the treatment assignment is independent of the potential outcomes and potential treatment-continuation statuses:
    \begin{equation*}
        X_i \perp \{Z_{i,T}(x,d), D_i(x)\} \mid Y_{i,0} \quad \text{for} \quad (x,d) \in \{0,1\}^2.
    \end{equation*}
    \item[(A5)] \textit{Exchangeability for treatment discontinuation}: Conditional on the treatment assignment and baseline measurement, the treatment-continuation status is independent of the potential outcomes:
    \begin{equation*}
        Z_{i,T}(x,d) \perp D_i \mid X_i = x, Y_{i,0} \quad \text{for} \quad (x,d) \in \{0,1\}^2.
    \end{equation*}
\end{enumerate}
These assumptions are somewhat stronger than those required under a formulation based solely on the one-argument potential outcome. However, given the randomized design and the simplified baseline-adjusted structure considered here, they remain reasonable for our analysis. In particular, exchangeability of treatment assignment follows from the randomized trial design, while the conditional exchangeability of treatment discontinuation reflects the assumption that, in this model, discontinuation behavior is adequately captured by observed participant characteristics, which are represented by the baseline measurement.

Under the hypothetical strategy, the estimand targets the population-level contrast in the endpoint under a hypothetical scenario in which treatment discontinuation had not occurred \citep{ICH2019E9R1}. Accordingly, the overall treatment effect for all randomized subjects under the hypothetical strategy can be expressed as $\mathbb{E} \left[ Z_{i, T} (1, 1) - Z_{i, T} (0, 1) \right]$. Assuming (A1)-(A5), we further define the conditional treatment effect, given the baseline measurement, under the hypothetical strategy as
\begin{equation}
    \begin{aligned}
        \mathbb{E} \left[ Z_{i,T}(1,1) - Z_{i,T}(0,1) \mid Y_{i,0} \right]
        &= \mathbb{E} \left[ Z_{i,T} \mid X_i = 1, D_i = 1, Y_{i,0} \right] - \mathbb{E} \left[ Z_{i,T} \mid X_i = 0, D_i = 1, Y_{i,0} \right] \\
        &= \mathbb{E} \left[ \mathbf{W}_i^\top \boldsymbol{\beta} \mid X_i = 1, Y_{i,0} \right] - \mathbb{E} \left[ \mathbf{W}_i^\top \boldsymbol{\beta} \mid X_i = 0, Y_{i,0} \right] \\
        &= \beta_X.
    \end{aligned}
\end{equation}
Therefore, the conditional treatment effect coincides with the overall treatment effect under the hypothetical strategy:
\begin{equation}
    \mathbb{E} \left[ Z_{i,T}(1,1) - Z_{i,T}(0,1) \right] = \mathbb{E}_{Y_{i,0}} \left[ \mathbb{E} \left[ Z_{i,T}(1,1) - Z_{i,T}(0,1) \mid Y_{i,0} \right] \right] = \beta_X.
\end{equation}
The equality between the treatment effect under the hypothetical strategy and the treatment coefficient $\beta_X$ holds within the proposed endpoint analysis model and requires correct specification of that model. As shown in Section \ref{subsec:infer}, $\beta_X$ can be estimated straightforwardly using maximum-likelihood inference.

Under the TP strategy, the estimand targets the population-level contrast in the endpoint for all randomized subjects, regardless of whether treatment discontinuation occurs \citep{ICH2019E9R1}. Accordingly, the overall treatment effect for all randomized subjects under the TP strategy can be expressed as $\mathbb{E} \left[Z_{i, T} (1, D_i(1)) - Z_{i, T} (0, D_i(0)) \right]$. Assuming (A1)-(A5), we further define the conditional treatment effect, given the baseline measurement, under the TP strategy as
\begin{equation}
    \begin{aligned}
        \mathbb{E} \left[ Z_{i,T}(1,D_i(1)) - Z_{i,T}(0,D_i(0)) \mid Y_{i,0} \right]
        &=
        \mathbb{E} \left[ Z_{i,T} \mid Y_{i,0}, X_i = 1 \right]
        - \mathbb{E} \left[ Z_{i,T} \mid Y_{i,0}, X_i = 0 \right] \\
        &=
        \mathbb{E} \left[ \mathbf{W}_i^\top \boldsymbol{\beta} \mid Y_{i,0}, X_i = 1 \right]
        - \mathbb{E} \left[ \mathbf{W}_i^\top \boldsymbol{\beta} \mid Y_{i,0}, X_i = 0 \right] \\
        &\qquad
        + \delta \left\{
        P(D_i = 0 \mid Y_{i,0}, X_i = 1)
        - P(D_i = 0 \mid Y_{i,0}, X_i = 0)
        \right\} \\
        &=
        \beta_X + \delta \cdot \Phi \left( \gamma_0 + Y_{i,0} \, ; \, \gamma_0 + Y_{i,0} + \gamma_X \right),
    \end{aligned}
\end{equation}
where $\Phi(a \,;\, b) = \Phi(b) - \Phi(a)$ and $\Phi(\cdot)$ is the cumulative distribution function of $N(0, 1)$. Note that this equality holds only if the endpoint model is correctly specified. The final term on the right-hand side is equal to $\delta$ multiplied by the difference in the conditional probabilities of treatment discontinuation between the treatment groups, given the baseline measurement $Y_{i,0}$. This term is equal to zero if treatment discontinuation is conditionally independent of $X_i$ given $Y_{i,0}$, that is, when $\gamma_X = 0$. In general, without specifying the distribution of $Y_{i,0}$, the overall treatment effect under the TP strategy is given by 
\begin{equation}
\beta_X^{\mathrm{TP}}
\equiv \mathbb{E} \left[ Z_{i,T}(1, D_i (1)) - Z_{i,T}(0, D_i (0)) \right] = \beta_X + \delta \cdot \mathbb{E}_{Y_{i,0}} \left[ \Phi \left( \gamma_0 + Y_{i,0} \,;\, \gamma_0 + Y_{i,0} + \gamma_X \right) \right],
\end{equation}
where $\mathbb{E}_{Y_{i,0}}[h(Y_{i,0})]$ denotes the expectation of $h(Y_{i,0})$ with respect to $Y_{i,0}$. Our formulation implies that evaluating $\beta_X^{\mathrm{TP}}$ depends on the distribution of the baseline measurement $Y_{i,0}$. For example, when $Y_{i,0} \sim N(\mu_0,\sigma_0^2)$, a more explicit expression for the overall treatment effect under the TP strategy is
\begin{equation}
\beta_X^{\mathrm{TP}} = \beta_X + \delta \cdot \Phi \left( \frac{\gamma_0+\mu_0}{\sqrt{1+\sigma_0^2}} \,;\, \frac{\gamma_0+\mu_0+\gamma_X}{\sqrt{1+\sigma_0^2}} \right) = \beta_X + \delta \cdot \left[ \Phi \left( \frac{\gamma_0+\mu_0+\gamma_X}{\sqrt{1+\sigma_0^2}} \right) - \Phi \left( \frac{\gamma_0+\mu_0}{\sqrt{1+\sigma_0^2}} \right) \right].
\end{equation}
The TP-strategy estimator $\beta_X^{\mathrm{TP}}$ can be decomposed into $\beta_X$ (hypothetical-strategy estimator) plus an additional term involving the treatment-arm difference in treatment discontinuation probabilities. This decomposition clarifies that the discrepancy between the two estimators arises from the extent to which treatment discontinuation both affects the endpoint and differs across treatment arms. In particular, the two estimators are identical when treatment discontinuation is conditionally independent of treatment assignment given baseline measurement. Our ML-based approach in Section \ref{subsec:infer} estimates $\beta_X^{\mathrm{TP}}$ by plugging the fitted endpoint-analysis and discontinuation models into the above expression and averaging over the empirical baseline distribution. 

\subsection{Estimation and Inference}\label{subsec:infer}

The ML approach can be used to estimate the parameters in our model by maximizing the observed log-likelihood, as derived in \ref{subsec:obsl}, the observed log-likelihood can be given as
\begin{equation}\label{main:obslik}
    \begin{aligned}
        \ell_{\text{obs}} \left(\mathbf{\beta}, \delta,  \mathbf{\gamma}, \pi, \sigma_{\epsilon}^2 \right)
        &= \sum_{i:D_{i}=1} \left\{ \log\phi\left(Z_{i,T} ; \mathbf{W}_{i}^\top\mathbf{\beta},\sigma_{\epsilon}^{2}\right) + \log\Phi\left(-\mathbf{W}_{i}^\top\mathbf{\gamma} \right) \right\} +\\
        &\quad \quad \quad \quad \sum_{i:D_{i}=0,Q_{i}=1} \left\{ \log\phi \left( Z_{i,T} ; \mathbf{W}_{i}^\top\mathbf{\beta} + \delta, \sigma_{\epsilon}^{2} \right) + \log(\pi) + \log\Phi \left( \mathbf{W}_{i}^\top\mathbf{\gamma} \right) \right\} +\\
        &\quad \quad \quad \quad \quad \quad \sum_{i:D_{i}=0,Q_{i}=0} \left\{ \log(1-\pi) + \log\Phi \left( \mathbf{W}_{i}^\top\mathbf{\gamma} \right) \right\}.
    \end{aligned}
\end{equation}
 By rearranging the terms, $\ell_{\text{obs}} \left(\mathbf{\beta}, \delta,  \mathbf{\gamma}, \pi, \sigma_{\epsilon}^2 \right)$ can be written as the summation of three components: The first component is 
\begin{equation}
    \sum_{i:D_{i}=1} \log\phi\left(Z_{i,T} ; \mathbf{W}_{i}^\top\mathbf{\beta},\sigma_{\epsilon}^{2}\right) + \sum_{i:D_{i}=0,Q_{i}=1} \log\phi \left(Z_{i,T} ; \mathbf{W}_{i}^\top\mathbf{\beta} + \delta, \sigma_{\epsilon}^{2} \right)
\end{equation}
which is independent of $\mathbf{\gamma}$ and $\pi$. The second component is 
\begin{equation}
    \sum_{i:D_{i}=1} \log\Phi\left(-\mathbf{W}_{i}^\top\mathbf{\gamma} \right) + \sum_{i:D_{i}=0} \log\Phi \left( \mathbf{W}_{i}^\top\mathbf{\gamma} \right)
\end{equation}
which is independent of $\mathbf{\beta}$, $\delta$, $\sigma_{\epsilon}^2$ $\pi$. The third component is
\begin{equation}
    \sum_{i:D_{i}=0,Q_{i}=1} \log(\pi) + \sum_{i:D_{i}=0,Q_{i}=0} \log(1-\pi)
\end{equation}
which is independent of $\mathbf{\beta}$, $\delta$, $\sigma_{\epsilon}^2$ and $\mathbf{\gamma}$. This implies that (a) $\mathbf{\beta}$, $\delta$ and $\sigma_{\epsilon}^2$; (b) $\mathbf{\gamma}$; and (c) $\pi$ can be separately estimated. For (a), we can follow the same approach as in the classic linear regression. For (b), any standard method for fitting generalized linear model can be used, such as the Newton-Raphson method. For (c), a closed-form estimator can be obtained. Since it is straightforward to compute these estimates, we refer the detailed derivation to \ref{subsec:mle}.

The derived ML estimators can be used to further estimate the treatment effect under each IE strategy in Section \ref{subsec:causal}. For simplicity, and consistent with Section \ref{subsec:causal}, we assume in the remainder of this section that $\mathbf{X}$ includes only treatment assignment as a single covariate. The overall treatment effect is $\beta_{X}$ under the hypothetical strategy that can be estimated by the corresponding ML estimator, $\hat{\beta}_{X}$, obtained from linear regression. Thus, the test statistic for $H_{0}: \beta_{X} = \beta_{0}$ is 
\begin{equation}
    \frac{\hat{\beta}_{X} - \beta_{0}}{\text{se}(\hat{\beta}_{X})} \sim t_{d},
\end{equation}
where $t_{d}$ denotes $t$-distribution with the degree of freedom $d=N-\sum_{i=1}^{N}\mathbf{1}\{D_{i}=0,Q_{i}=0\}-4$ and $\text{se}(\hat{\beta}_{X})$ is the standard error of $\hat{\beta}_{X}$ as in the linear regression, see \ref{subsec:mle}.

Recall that the overall treatment effect is $\beta_X^{\mathrm{TP}} = \beta_X + \delta \cdot \mathbb{E}_{Y_{i,0}} \left[ \Phi \left( \gamma_0 + Y_{i,0} \,;\, \gamma_0 + Y_{i,0} + \gamma_X \right) \right]$ under the TP strategy. By plugging in the corresponding ML estimators of the parameters in the TP-strategy estimator and replacing the expectation with the sample mean, we obtain the estimator for $\beta_{X}^{\text{TP}}$ as 
\begin{equation}
    \hat{\beta}^{\text{TP}}_{X} = \hat{\beta}_{X} + \hat{\delta}\cdot \frac{1}{N} \sum_{i=1}^{N} \Phi \left( \hat{\gamma}_{0} + Y_{i, 0} ; \hat{\gamma}_0 + Y_{i, 0} + \hat{\gamma}_{X} \right).
\end{equation}
Since the estimator is a nonlinear function of the model parameters, it is challenging to derive a closed-form expression for its standard error. In such cases, resampling methods such as the bootstrap can be used to estimate the standard error and perform statistical inference \cite{Hill2006}. The proposed bootstrap procedure for obtaining $B$ bootstrap samples is as follows: for each $b = 1, \ldots, B$,
\begin{enumerate}
    \item Generate $\check{\eta}_{i}^{(b)} \stackrel{i.i.d.}{\sim} N(0,1)$ and compute $$\check{D}_{i}^{(b)}=\mathbf{1}\left\{ \hat{\gamma}_{0} + Y_{i,0} + \hat{\gamma}_{X}X_{i} + \check{\eta}_{i}^{(b)} < 0 \right\} \quad \text{for} \quad i=1,\dots,N.$$ Then, calculate $\hat{\gamma}_{0}^{(b)}$ and $\hat{\gamma}_{X}^{(b)}$ using $\{ X_{i},Y_{i,0},\check{D}_{i}^{(b)}\}_{i=1}^{N}$.
    \item Without loss of generality, suppose the first $M$ observations corresponding to either completers or RDs, after appropriate rearrangement. For each $i=1,\ldots,M$, compute the bootstrapped response as $$Y_{i,T}^{(b)} =  \hat{\beta}_{0} + (1+\hat{\beta}_{\text{base}}) Y_{i,0} + \hat{\beta}_{X}X_{i} + \mathbf{1}_{[D_i =0]} \hat{\delta} + \check{\epsilon}_{i},$$ where $\check{\epsilon}_{i}=Z_{i,T} - \hat{\beta}_{0} - \hat{\beta}_{\text{base}} Y_{i,0} - \hat{\beta}_{X} X_{i} - \mathbf{1}_{[D_i =0]} \hat{\delta}$ is a bootstrap residual resampled with replacement from the original residuals. Then, calculate $\hat{\beta}_{X}^{(b)}$ and $\hat{\delta}^{(b)}$ using $\{X_{i},D_{i},Q_{i},Z_{i,T}^{(b)}\}_{i=1}^{M}$ where $Z_{i,T}^{(b)}=Y_{i,T}^{(b)}-Y_{i,0}$.
    \item The $b$th bootstrap estimator for $\beta_{X}^{\text{TP}}$ is calculated as follows: $$\hat{\beta}^{\text{TP}, (b)}_{X} = \hat{\beta}_{X}^{(b)} + \hat{\delta}^{(b)} \frac{1}{N} \sum_{i=1}^{N} \Phi \left( \hat{\gamma}_{0}^{(b)} + Y_{i, 0} ; \hat{\gamma}_{0}^{(b)} + Y_{i, 0} + \hat{\gamma}_{X}^{(b)} \right).$$
    
\end{enumerate}
Based on these bootstrap replicates, the standard error can be estimated by the sample standard deviation of $\hat{\beta}^{\text{TP}, (1)}_{X}, \ldots, \hat{\beta}^{\text{TP}, (B)}_{X}$ . In addition, the $100(1-\alpha)\%$ bootstrap confidence interval (CI) \citep{Davison1997, Hesterberg2015} for $\beta_{X}^{\text{TP}}$ can be given as
\begin{equation}
    \left( 2 \hat{\beta}^{\text{TP}}_{X} - q_{1-\alpha/2,B},\quad 2 \hat{\beta}^{\text{TP}}_{X} - q_{\alpha/2,B} \right),
\end{equation}
where $q_{\alpha,B}$ is the empirical $\alpha$-quantile of $\hat{\beta}^{\text{TP}, (1)}_{X}, \ldots , \hat{\beta}^{\text{TP}, (B)}_{X}$.

\section{Simulation studies}\label{sec:simul}

We conduct simulation studies to assess the performance of our approach. For comparison, three common imputation methods are considered under the TP strategy: RTB imputation \citep{Qu2022}, washout imputation, and RD imputation \citep{Wang2022}, each combined with standard ANCOVA for efficacy analysis. For each method, we generate 1000 imputed datasets and combine resulting outcomes using Rubin's rule \citep{Rubin1996}. Notably, RD imputation assumes that the endpoint values in RDs are representative of those that would have been observed in unretrieved dropouts, conditional on observed covariates. While this assumption is the same as that underlying our approach, our proposed method differs in that it explicitly and jointly models the endpoint and treatment/study discontinuation processes, and estimates the treatment effect in a model-based manner. We refer to \ref{subsec:cmpt} for details on each competing method.

For comprehensive simulations, we vary (a) magnitudes of on-treatment and post-treatment-discontinuation treatment effects; (b) treatment discontinuation rate in probit models. For (a), we try (i) $\beta_X = -10$ and $\delta = 5$, representing an efficacious experimental drug for which part of the treatment effect persists after treatment discontinuation; (ii) $\beta_X = -10$ and $\delta = 10$, representing an efficacious drug for which the treatment effect does not persist after treatment discontinuation, so that the entire treatment effect is lost; (iii) $\beta_X = 0$ and $\delta = 0$, representing an inefficacious drug. For (b), $\gamma_X=\pm0.25$ are considered to induce differential treatment-discontinuation rates between the two arms where the treatment discontinuation rate is higher in the placebo (or experimental arm) for the negative (or positive) value. For the other configurations, we set $\beta_{0}=\beta_{\text{base}}=0$, $\gamma_0=-0.75$, $\pi = 0.5$, $\sigma_\epsilon^2=20^2$ and simulate $Y_{i,0}\stackrel{i.i.d.}{\sim}N(180,20^2)$. In the numerical implementation, the baseline term entering the probit discontinuation model was represented on a standardized scale; this does not change the model formulation or the interpretation of the resulting treatment effect expressions. Each simulation scenario is independently repeated for $N_{\text{sim}} = 5000$ times with sample size $N = 200$. The results with $N=100$ are presented in Tables \ref{tab:simout3} and \ref{tab:simout4} under \ref{subsec:simout}. We assess the performance of each method using five metrics: average bias, root mean squared error (RMSE), empirical rejection rate for the nominal two-sided test of $H_{0}:\beta_{X}^{\text{TP}}=0$ at the 5\% significance level, and empirical coverage probability and average length of the 95\% CI for \(\beta_X^{\text{TP}}\) over 5000 simulation runs. In the efficacious Scenarios 1 and 2, the rejection rate may be interpreted as empirical power, whereas in the inefficacious Scenario 3 it corresponds to empirical type I error.

Table \ref{tab:simout1} shows the simulation results for the three combinations of \((\beta_X, \delta)\) when the placebo arm has a higher rate of treatment discontinuation with \(\gamma_X = -0.25\) compared to the experimental arm. Specifically, around 30\% and 24\% are the treatment discontinuation rates in the placebo and experimental arms, respectively, with about half of those stopping the study entirely. As expected from the decomposition in Section 2.3, the treatment effect under the TP strategy is larger in absolute value than the corresponding treatment effect under the hypothetical strategy, because discontinuation is less frequent in the experimental arm and discontinuation attenuates efficacy. In the efficacious scenarios, RTB and washout imputation exhibit substantial positive bias, whereas RD imputation and the proposed method remain close to unbiased. RTB imputation achieves smaller RMSE than the proposed method; however, this apparent advantage is accompanied by appreciable bias. Relative to RD imputation, the proposed method preserves similarly small bias while achieving lower RMSE, higher rejection rates when the drug is efficacious, rejection rates close to 0.05 when the drug is inefficacious, and empirical coverage closer to the nominal 95\% level. In the inefficacious scenario, all methods have small bias, but the proposed method provides the rejection rate and coverage closest to the nominal levels, with CI length comparable to the shortest.

The sizable bias of RTB and washout imputation in Scenario 2 of Table \ref{tab:simout1} may seem counterintuitive, since neither method assumes a persisting post-discontinuation treatment effect. However, RTB and washout imputation do not account for the background rate of treatment discontinuation, irrespective of treatment assignment. In contrast, RD imputation and the proposed method incorporate post-discontinuation information and therefore better accommodate discontinuation arising for reasons unrelated to the experimental drug, as represented by the placebo-arm discontinuation rate and by real-world factors such as personal circumstances. This distinction is important for accurate estimation of the treatment effect under the TP strategy.

Table \ref{tab:simout2} presents results for the three treatment effect scenarios with a higher treatment discontinuation rate in the experimental arm than the placebo arm, which is opposite to the setup in Table \ref{tab:simout1}. Specifically, the treatment discontinuation rates are around 30\% and 36\% in the placebo and experimental arms, respectively, with about half of those stopping the study entirely. As in Table \ref{tab:simout1}, RTB and washout imputation again show substantial positive bias in the efficacious scenarios, whereas RD imputation and the proposed method remain close to the target. Although RTB imputation yields a smaller RMSE, this reflects a bias-variance trade-off rather than closer alignment with the TP strategy. Compared with RD imputation, the proposed method retains similar bias, reduces RMSE, and provides better-calibrated inference, with higher rejection rates when the drug is efficacious, rejection rates close to 0.05 when the drug is inefficacious, and empirical coverage closer to the nominal 95\% level. In the inefficacious scenario, washout imputation attains the smallest RMSE but is markedly conservative, whereas the proposed method remains better calibrated in terms of rejection rate and coverage.

Since RD imputation relies heavily on the availability of observed RD data, its performance may deteriorate when the retrieval probability $\pi$ is low. Motivated by this, we conducted additional simulations with lower retrieval probabilities, $\pi=0.3$ and $\pi=0.1$. The results are reported in Tables \ref{tab:simout5}-\ref{tab:simout8} in \ref{subsec:simout}. As $\pi$ decreases, RD imputation shows substantial RMSE inflation, declining rejection rate in the efficacious scenarios, and pronounced under-coverage together with inflated rejection rate in the inefficacious scenario. By contrast, the proposed method retains small bias and near-nominal inference across scenarios, even though its CIs may be somewhat longer than those from RTB or washout imputation in the settings with smaller $\pi$. For example, when $\pi=0.1$ and $N = 200$, the empirical coverage of RD imputation drops to approximately 70-79\% across scenarios, whereas the proposed method remains close to 95\%. These results suggest that the main advantage of the proposed method is robustness and inferential calibration when the amount of observed RD data is limited.

\section{Illustrative Application}\label{sec:realdata}

We illustrate the proposed method using an application inspired by a publicly available dataset from an antidepressant clinical trial \cite{Mallinckrodt2014}. The endpoint is the change from baseline to Week 6 on the Hamilton 17-item rating scale for depression (HAMD17). The analysis focuses on the difference in mean change between treatment arms. Among the available time points (Weeks 1, 2, 4, and 6), only baseline and Week 6 data are used in the analysis, because Week 8 data are not publicly available to preserve the confidentiality of the original trial results. The investigational antidepressant evaluated in this trial has since received regulatory approval and is now widely prescribed \cite{Detke2004,Goldstein2004}. The dataset has previously been used to evaluate statistical methods and to illustrate attributes of strategies described in ICH E9(R1) \cite{Jin2020}, and is available at \url{https://www.lshtm.ac.uk/research/centres-projects-groups/missing-data}. 

The primary estimand, defined in accordance with the five attributes in ICH E9(R1), is as follows: the population comprises all randomized subjects in the dataset (84 assigned to experimental drug and 88 assigned to placebo); the variable is the change from baseline to Week 6 in HAMD17 total score; the IE of interest is treatment discontinuation, which is handled using a TP strategy, such that the outcome is defined regardless of whether treatment discontinuation occurs; the population-level summary is the difference between treatment groups in the mean change from baseline to Week 6 in HAMD17 total score; accordingly, the estimand is the mean treatment difference between the randomized treatment groups in change from baseline to Week 6 in HAMD17 total score among all randomized subjects, regardless of treatment discontinuation after randomization.

\begin{figure}[t]
    \centering
    \includegraphics[width=1\textwidth]{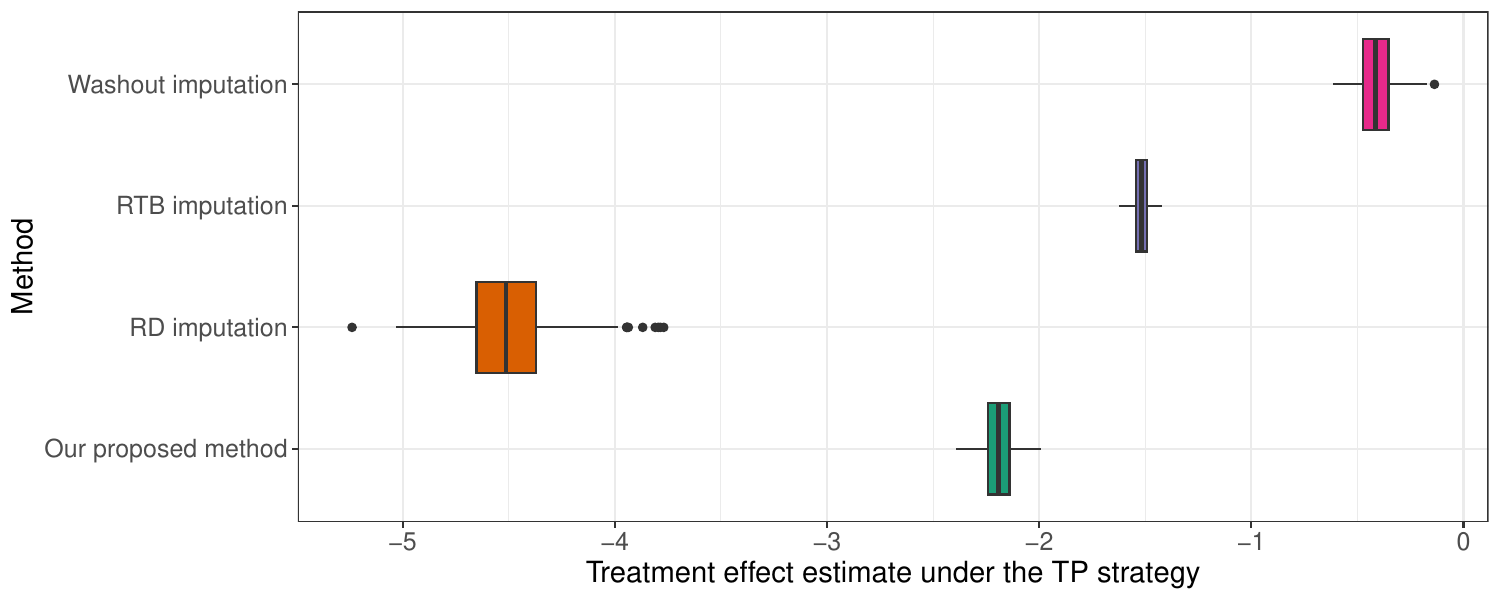}
    \caption{Treatment effect estimates under the TP strategy based on 500 synthetic HAMD17 datasets generated from the original trial data under partial treatment effect persistence after discontinuation}
    \label{fig:application}
\end{figure}

The publicly available dataset does not contain RD information. To assess the performance of the proposed method in the presence of RDs, we introduce a structured synthetic RD generation procedure using the two-step approach\cite{Wang2023}, while preserving the baseline and response characteristics of the original data. In the selection step, a subset of completers is reclassified as treatment-discontinued subjects. Subjects in the experimental arm are assumed to discontinue treatment due to adverse events, with $k\%$ selected from the top $(k+10)\%$ of performers based on the change from baseline in HAMD17 at Week 6. In contrast, subjects in the placebo arm are assumed to discontinue due to lack of efficacy, with $k\%$ selected from the bottom $(k+10)\%$ of performers, which was proposed to approximate plausible clinical trial conditions \cite{Wang2023}. In the replacement step, the change from baseline in HAMD17 for the selected subjects is multiplied by $0.5$ to reflect partial persistence of the treatment effect after discontinuation. The no-persistence scenario, obtained by multiplying by $0$, is reported in \ref{subsec:appout_add}. This procedure generates trial-inspired synthetic datasets with RD structures, enabling evaluation of the proposed method under a setting motivated by a real clinical trial. Notably, it creates observed RDs from the tails of the Week-6 response distribution within each treatment arm, a feature that is useful for interpreting the behavior of RD imputation in Figure \ref{fig:application}.

Figure \ref{fig:application} compares treatment effect estimates under the TP strategy across four methods: the proposed method, RTB imputation, washout imputation, and RD imputation. The ANCOVA analysis is repeated using 500 synthetic HAMD17 datasets with RDs generated according to the procedure described above, with $k = 10$ assumed. A notable feature is that RD imputation yields substantially more negative estimates than the other methods. This divergence is plausible given the way the synthetic RDs were generated. In the experimental arm, the selected RDs are drawn from subjects with relatively strong observed Week-6 improvement, whereas in the placebo arm they are drawn from subjects with relatively weak observed improvement. Even after the replacement step using a multiplier of 0.5, the resulting observed RDs remain a selected subset with more favorable outcomes in the experimental arm and less favorable outcomes in the placebo arm than would typically be expected among all discontinuers. Since RD imputation uses the observed RD distribution directly to impute missing data for unretrieved discontinuers, it is particularly sensitive to this lack of representativeness and may therefore amplify the between-arm contrast. By contrast, the proposed method uses RD information through a joint model for the endpoint and discontinuation processes rather than through direct imputation driven by the observed RD data, which likely explains its more moderate estimates and substantially smaller variability. When there is no persisting treatment effect, as shown in Figure \ref{fig:application_add}, the proposed method is close to RD imputation in the location of the estimates, although RD imputation still exhibits greater variability than the proposed method and the other competing methods.

\section{Discussion}\label{sec:concl}

We developed a likelihood-based modeling framework for continuous endpoints in randomized trials with treatment discontinuation and RDs. The approach combines an ANCOVA model for the endpoint analysis with an explicit model for the treatment-discontinuation process, enabling transparent formulation of treatment effects for estimands defined using the hypothetical and TP strategies for the IE of treatment discontinuation. Estimation proceeds by maximizing the observed-data likelihood, and the TP-strategy treatment effect estimator is obtained via a plug-in expression that depends on both the fitted endpoint and discontinuation components.

A key practical motivation is that RD observations provide post-discontinuation data, which may differ systematically from on-treatment data. Methods that either pool completers and RDs without distinction or rely directly on observed RDs for imputing missing outcomes may obscure the relationship between modeling assumptions and the target estimand under the TP strategy. Across the main simulation scenarios, the proposed method consistently maintained small bias and better inferential calibration than commonly used imputation-based approaches. Relative to RD imputation, it generally preserved similarly small bias while reducing RMSE and producing rejection rates and coverages closer to the nominal levels. In additional numerical experiments with lower retrieval probabilities, RD imputation deteriorated markedly as the amount of observed RD data decreased, whereas the proposed method remained comparatively stable and retained near-nominal inference. The illustrative application showed a similar pattern: RD imputation produced substantially more negative, and possibly overestimated, treatment effect estimates, plausibly because the structured RD generation procedure relies on observed RDs from the tails of the response distribution, whereas the proposed method yielded more stable estimates.

The proposed approach is model-based and therefore has limitations that merit emphasis. First, it adopts an ANCOVA formulation to ease interpretation. In practice, discontinuation may depend on intermediate data observed during follow-up. We therefore provide a longitudinal extension in the Supplementary Material that allows history-dependent discontinuation and outlines likelihood-based estimation via an expectation-maximization-type procedure. Second, interpretation under the hypothetical strategy relies on correct specification of the post-discontinuation component (parameterized by $\delta$). Misspecification of the discontinuation-associated shift may affect the ability to recover an on-treatment contrast from models that incorporate post-discontinuation responses. Third, the retrieval mechanism is parameterized in a parsimonious manner. Although this choice improves stability and interpretability, more flexible retrieval models (e.g., covariate-dependent retrieval) may be needed in some applications, and their incorporation would require careful development of valid inference.

Several methodological extensions are promising. The discontinuation model can be generalized beyond probit regression (e.g., logistic regression), and the endpoint analysis model can incorporate treatment-specific post-discontinuation shifts through interaction terms, as described in the Supplementary Material. In addition, relaxing the independence between the regression error in the endpoint analysis model and the discontinuation process would allow the framework to accommodate settings in which discontinuation depends on unobserved determinants of the endpoint, analogous to non-ignorable discontinuation mechanisms. Developing principled sensitivity analyses within this joint modeling framework is an important direction for future research.

Finally, our illustrative application was based on a publicly available antidepressant trial dataset that does not contain RD information. Following prior work\cite{Wang2023}, we used a structured RD generation procedure to create synthetic RD data while preserving key baseline and response characteristics of the original data. This example is intended to demonstrate how the proposed estimator behaves under a realistic, trial-motivated data structure rather than to reproduce the original regulatory analysis. Broader evaluation on real datasets with observed RD data remains an important goal for future research as such data become available.

\section*{Acknowledgments}
At the time of this study, Myeongjong Kang was employed by Merck $\&$ Co$.$, Inc. The views expressed in this manuscript are solely those of the authors and do not necessarily reflect the views of Merck $\&$ Co$.$, Inc$.$ or its affiliates. Merck $\&$ Co$.$, Inc$.$ makes no representations or warranties as to the accuracy or reliability of the information presented herein. For an earlier version of computing in Section \ref{sec:simul}, Sangyoon Yi acknowledges the High Performance Computing Center at Oklahoma State University supported in part through the National Science Foundation grant OAC-1531128.

\section*{Conflict of Interests}
The authors declare no conflicts of interest.

\section*{Data Availability Statement}
The HAMD17 data in Section \ref{sec:realdata} is available at \url{https://www.lshtm.ac.uk/research/centres-projects-groups/missing-data}.

\nocite{*}%
\bibliography{ref}%

\clearpage
\renewcommand{\arraystretch}{1.1}
\begin{table}[t]
\centering
\caption{Performance of each method under three scenarios with higher treatment discontinuation rate in the placebo arm ($\gamma_X = -0.25$) and sample size $N=200$}
\vspace{1em}
\begin{tabular}{@{}lcccccc@{}}
\toprule \toprule
\multicolumn{1}{c}{\multirow{2}{*}{Method}} &
  \multirow{2}{*}{$\beta_X^{\text{TP}}$} &
  \multirow{2}{*}{Bias} &
  \multirow{2}{*}{RMSE} &
  \multirow{2}{*}{Rejection Rate} &
  \multicolumn{2}{c}{95$\%$ CI} \\ \cmidrule(l){6-7}
\multicolumn{1}{c}{} & & & & & Coverage ($\%$) & Length \\ \midrule

\multicolumn{7}{l}{Scenario 1: Efficacious drug with persisting effect ($\beta_X = -10$ and $\delta = 5$)} \\ \midrule
RTB imputation       & \multirow{4}{*}{-10.291} & \phantom{ }1.341 & 2.999 & 0.869 & 95.22 & 11.910 \\
Washout imputation   &                          & \phantom{ }2.679 & 3.711 & 0.691 & 91.66 & 12.595 \\
RD imputation        &                          & \phantom{ }0.022 & 3.279 & 0.903 & 92.78 & 11.907 \\
Our method           &                          & \phantom{ }0.005 & 3.088 & 0.916 & 94.28 & 11.897 \\ \midrule

\multicolumn{7}{l}{Scenario 2: Efficacious drug without persisting effect ($\beta_X = -10$ and $\delta = 10$)} \\ \midrule
RTB imputation       & \multirow{4}{*}{-10.582} & \phantom{ }1.474 & 3.048 & 0.881 & 95.04 & 12.012 \\
Washout imputation   &                          & \phantom{ }2.978 & 3.921 & 0.701 & 90.02 & 12.581 \\
RD imputation        &                          & -0.020 & 3.285 & 0.924 & 93.00 & 12.042 \\
Our method           &                          & -0.016 & 3.087 & 0.931 & 94.52 & 12.014 \\ \midrule

\multicolumn{7}{l}{Scenario 3: Inefficacious drug ($\beta_X = 0$ and $\delta = 0$)} \\ \midrule
RTB imputation       & \multirow{4}{*}{0.000}   & -0.027 & 2.556 & 0.020 & 98.02 & 11.815 \\
Washout imputation   &                         & -0.022 & 2.479 & 0.012 & 98.84 & 12.490 \\
RD imputation        &                         & -0.037 & 3.152 & 0.062 & 93.76 & 11.861 \\
Our method           &                         & -0.024 & 2.969 & 0.045 & 95.40 & 11.841 \\ \bottomrule \bottomrule
\end{tabular}
\label{tab:simout1}
\end{table}

\clearpage
\renewcommand{\arraystretch}{1.1}
\begin{table}[t]
\centering
\caption{Performance of each method under three scenarios with higher treatment discontinuation rate in the experimental arm ($\gamma_X = 0.25$) and sample size $N=200$}
\vspace{1em}
\begin{tabular}{@{}lcccccc@{}}
\toprule \toprule
\multicolumn{1}{c}{\multirow{2}{*}{Method}} &
  \multirow{2}{*}{$\beta_X^{\text{TP}}$} &
  \multirow{2}{*}{Bias} &
  \multirow{2}{*}{RMSE} &
  \multirow{2}{*}{Rejection Rate} &
  \multicolumn{2}{c}{95$\%$ CI} \\ \cmidrule(l){6-7}
\multicolumn{1}{c}{} & & & & & Coverage ($\%$) & Length \\ \midrule

\multicolumn{7}{l}{Scenario 1: Efficacious drug with persisting effect ($\beta_X = -10$ and $\delta = 5$)} \\ \midrule
RTB imputation       & \multirow{4}{*}{-9.681} & \phantom{ }1.629 & 3.099 & 0.774 & 94.82 & 12.074 \\
Washout imputation   &                          & \phantom{ }3.279 & 3.987 & 0.481 & 91.08 & 12.887 \\
RD imputation        &                          & -0.025 & 3.346 & 0.861 & 92.18 & 12.055 \\
Our method           &                          & -0.018 & 3.133 & 0.874 & 94.18 & 12.084 \\ \midrule

\multicolumn{7}{l}{Scenario 2: Efficacious drug without persisting effect ($\beta_X = -10$ and $\delta = 10$)} \\ \midrule
RTB imputation       & \multirow{4}{*}{-9.361} & \phantom{ }1.484 & 3.016 &  0.747 & 95.90 & 12.225 \\
Washout imputation   &                          & \phantom{ }2.985 & 3.739 & 0.484 & 93.78 & 12.918 \\
RD imputation        &                          & -0.002 & 3.342 & 0.836 & 92.68 & 12.224 \\
Our method           &                          & -0.004 & 3.146 & 0.844 & 94.56 & 12.257 \\ \midrule

\multicolumn{7}{l}{Scenario 3: Inefficacious drug ($\beta_X = 0$ and $\delta = 0$)} \\ \midrule
RTB imputation       & \multirow{4}{*}{0.000}   & -0.054 & 2.581 &  0.022 & 97.78 & 11.976 \\
Washout imputation   &                         &  -0.038 & 2.217 &  0.003 & 99.68 & 12.779 \\
RD imputation        &                         &  -0.053 & 3.302 & 0.074 & 92.56 & 12.019 \\
Our method           &                         &  -0.068 & 3.104 & 0.057 & 94.04 & 12.053 \\ \bottomrule \bottomrule
\end{tabular}
\label{tab:simout2}
\end{table}

\clearpage
\setcounter{section}{0}
\setcounter{equation}{0}
\setcounter{figure}{0}
\setcounter{table}{0}

\makeatletter
\renewcommand{\thesection}{S\arabic{section}}
\renewcommand{\theequation}{S\arabic{equation}}
\renewcommand{\thetable}{S\arabic{table}}
\renewcommand{\thefigure}{S\arabic{figure}}
\renewcommand{\thefootnote}{\fnsymbol{footnote}}

\begin{center}
{\null
    {
        \Large Supplementary Material for ``Estimation of Treatment Effect in
    }\\
\vspace{0.5cm}
    {
        \Large Clinical Trials of Continuous Endpoints with Retrieved Dropouts" by
    }\\
\vspace{1cm}
    {
        \large Myeongjong Kang\footnote{mkangstat@gmail.com}$^{1}$ and Sangyoon Yi\footnote{sayi@okstate.edu}$^{\dagger2}$
    }\\
\vspace{0.75cm}
    {
        \large {$^1$}Merck \& Co., Inc., Rahway, New Jersey, USA
    }\\
\vspace{0.5cm}
    {
        \large {$^2$}Department of Statistics, Oklahoma State University, Stillwater, Oklahoma, USA
    } 
\vspace{1cm}}
\end{center}

\section{A unified expression for different endpoints}\label{subsec:unif}

In modern longitudinal clinical trials, three commonly used types of endpoints are the response itself ($Y_{i,T}$), change from baseline ($Y_{i,T} - Y_{i,0}$), and percent change from baseline ($100 \times (Y_{i,T} - Y_{i,0}) / Y_{i,0}$). These forms can be expressed using a unified formula: 
\begin{equation}
    Z_{i, T} = \omega_i (Y_{i,T} - \nu Y_{i,0})
\end{equation}
where $\nu$ and $\omega_i$ are shifting and scaling that define the endpoint types: 
\begin{itemize}
    \item For the response endpoint, $\omega_i = 1$ and $\nu = 0$. 
    \item For the change-from-baseline endpoint, $\omega_i = 1$ and $\nu = 1$.
    \item For the percent-change-from-baseline endpoint, $\omega_i = 100 / Y_{i,0}$ and $\nu = 1$. 
\end{itemize}
Our proposed method, including model formulation and estimation, can be easily applied for such general form of $Z_{i, T}$.

\section{Extension of the proposed method to longitudinal analysis}\label{subsec:extension}

The main manuscript focuses on an analysis of covariance (ANCOVA)-based, baseline-adjusted analysis at the final visit. To address settings in which treatment discontinuation may depend on intermediate data observed during follow-up, we extend the framework in this appendix to repeated measurements. We assume monotone treatment discontinuation and monotone study discontinuation. 

For subject $i=1,\ldots,N$ and visits $t=0,1,\ldots,T$, let $Y_{it}$ denote the continuous clinical response, where $t=0$ corresponds to baseline, and define
\[
Z_{it}=Y_{it}-Y_{i0} \quad \text{for} \quad t=1,\ldots,T.
\]
Let $\mathbf Z_i=(Z_{i1},\ldots,Z_{iT})^\top$. Let $X_i\in\{0,1\}$ denote treatment assignment and let $\mathbf C_i$ denote additional baseline covariates.

For $t=1,\ldots,T$, define the time-varying treatment status
\[
D_{it}=
\begin{cases}
1, & \text{if subject } i \text{ remains on treatment through visit } t,\\
0, & \text{otherwise,}
\end{cases}
\]
with $D_{i0}=1$ and $D_{it}\le D_{i,t-1}$. Note that the endpoint-level treatment continuation indicator in the main manuscript is recovered as $D_i=D_{iT}$. Among subjects with $D_i=0$, let
\[
Q_i=
\begin{cases}
1, & \text{if the final-visit response } Y_{iT} \text{ is retrieved},\\
0, & \text{otherwise.}
\end{cases}
\]
Thus, completers, retrieved dropouts (RDs), and lost-to-follow-up subjects (LTFUs) correspond to $D_i=1$, $(D_i,Q_i)=(0,1)$, and $(D_i,Q_i)=(0,0)$, respectively. To keep the formulation aligned with the main text, the retrieval indicator is retained only for the final visit, although a visit-specific retrieval process could also be introduced.

Let $\mathcal H_{i,t-1}$ denote the observed history available just prior to visit $t$, for example,
\[
\mathcal H_{i,t-1}=(Y_{i0},\mathbf C_i,X_i,Z_{i1},\ldots,Z_{i,t-1}),
\]
or a lower-dimensional summary thereof. We model the visit-specific discontinuation among subjects still at risk by
\[
\lambda_{it}(\mathcal H_{i,t-1})
\equiv
P(D_{it}=0 \mid D_{i,t-1}=1,\mathcal H_{i,t-1})
=
\Phi(\mathbf V_{it}^\top \boldsymbol\gamma_t),
\qquad t=1,\ldots,T,
\]
where $\mathbf V_{it}$ may include baseline covariates, treatment assignment, and functions of the observed outcome history. Equivalently,
\[
P(D_{it}=1 \mid D_{i,t-1}=1,\mathcal H_{i,t-1})
=
\Phi(-\mathbf V_{it}^\top \boldsymbol\gamma_t).
\]
Among subjects with $D_i=0$, we retain the endpoint-level retrieval model
\[
Q_i \mid (D_i=0) \sim \mathrm{Bernoulli}(\pi).
\]
We assume that for subjects with $D_i = 1$, $P(Q_i=1\mid D_i=1)=1$.

To incorporate longitudinal outcomes, we use the multivariate normal model
\[
\mathbf Z_i \mid Y_{i0},X_i,\mathbf C_i,\mathbf D_i
\sim
N_T(\boldsymbol\mu_i,\boldsymbol\Sigma),
\]
where $\mathbf D_i=(D_{i1},\ldots,D_{iT})^\top$, $\boldsymbol\Sigma$ is an unstructured (or parsimonious) covariance matrix, and
\[
\mu_{it}
=
\mathbf W_{it}^\top \boldsymbol\alpha_t
+
\beta_t X_i
+
\delta_t \, \mathbf{1}(D_{it}=0) \quad \text{for} \quad t=1,\ldots,T.
\]
Here $\mathbf{W}_{it}$ contains fully observed covariates such as an intercept, $Y_{i0}$, visit indicators, and $\mathbf C_i$. The parameter $\delta_t$ represents the mean shift at visit $t$ after treatment discontinuation. To mirror the main-manuscript formulation more closely, one may impose $\delta_t=0$ for $t<T$ and retain only a final-visit shift $\delta_T$.

The complete-data likelihood corresponding to the longitudinal extension is
\[
\begin{aligned}
L_c(\boldsymbol\theta)
=
&\prod_{i=1}^N
\Bigg[
f_{\mathrm{MVN}}
\!\left(
\mathbf Z_i
\mid
Y_{i0},X_i,\mathbf C_i,\mathbf D_i;
\boldsymbol\alpha,\boldsymbol\beta,\boldsymbol\delta,\boldsymbol\Sigma
\right) \times \\
&\prod_{t=1}^T
\lambda_{it}(\mathcal H_{i,t-1})^{\mathbf{1}(D_{i,t-1}=1,D_{it}=0)}
\left\{1-\lambda_{it}(\mathcal H_{i,t-1})\right\}^{\mathbf{1}(D_{i,t-1}=1,D_{it}=1)}
\Bigg] \times \\
&\pi^{\mathbf{1}(D_i=0,Q_i=1)}
(1-\pi)^{\mathbf{1}(D_i=0,Q_i=0)},    
\end{aligned}
\]
where $f_{\mathrm{MVN}}(\cdot)$ denotes the multivariate normal density with the mean vector of $\boldsymbol{\mu}_i$ and the covariance matrix of $\boldsymbol\Sigma$. The observed-data likelihood is obtained by integrating the complete-data likelihood over unobserved components of $\mathbf Z_i$. In particular, when only the final-visit response may be missing, this integration is one-dimensional for LTFUs.

An expectation-conditional-maximization (ECM) algorithm can be used for estimation. In the E-step, compute the conditional moments of the missing components of $\mathbf Z_i$ given the observed data under the current parameter values using conditional expectation formula for the multivariate normal distribution. In the first CM-step, update $(\boldsymbol\alpha, \boldsymbol\beta, \boldsymbol\delta, \boldsymbol\Sigma)$ by maximizing the expected observed-data log-likelihood. In the second CM-step, update the discontinuation parameters $\{\boldsymbol\gamma_t\}_{t=1}^T$ using visit-specific probit regressions among subjects with $D_{i,t-1}=1$; when $\mathbf V_{it}$ contains lagged responses that are partially unobserved, the corresponding conditional expectations from the E-step can be used within an ECM update. In the third CM-step, update $\pi$ by the following closed-form estimator:
\[
\widehat{\pi}
=
\frac{\#\{i:(D_i,Q_i)=(0,1)\}}
{\#\{i:D_i=0\}}.
\]

\section{Extension of the proposed method to treatment-specific post-discontinuation effects}\label{subsec:genmod}

The main-manuscript model assumes a common final-visit mean shift after treatment discontinuation. To allow the post-discontinuation effect to differ by treatment arm, we extend the endpoint model to
\[
Z_{i,T}
=
\mathbf{W}_i^\top\boldsymbol\beta
+
\mathbf{1}(D_i=0)\,(\delta_0+\delta_X X_i)
+
\epsilon_i,
\qquad
\epsilon_i \stackrel{\mathrm{i.i.d.}}{\sim} N(0,\sigma_\epsilon^2),
\]
where $X_i\in\{0,1\}$ denotes treatment assignment. Here $\delta_0$ represents the post-discontinuation mean shift in the placebo arm, whereas $\delta_X$ represents the additional post-discontinuation shift in the experimental arm. Setting $\delta_X=0$ recovers the model in the main manuscript.

The (dis)continuation and retrieval models remain $D_i= \mathbf{1} (\mathbf{W}_i^\top\boldsymbol\gamma+\eta_i<0)$ for $\eta_i \stackrel{\mathrm{i.i.d.}}{\sim} N(0,1)$, $Q_i\mid(D_i=0)\sim \mathrm{Bernoulli}(\pi)$, and $P(Q_i=1\mid D_i=1)=1$. Under this extension, the observed-data log-likelihood becomes
{\small
\begin{align*}
\ell_{\mathrm{obs}}(\boldsymbol\beta,\delta_0,\delta_X,\boldsymbol\gamma,\pi,\sigma_\epsilon^2)
=&
\sum_{i:D_i=1}
\left[
\log \phi\!\left(Z_{i,T};\mathbf W_i^\top\boldsymbol\beta,\sigma_\epsilon^2\right)
+
\log \Phi(-\mathbf W_i^\top\boldsymbol\gamma)
\right] \\
+
&\sum_{i:D_i=0,\ Q_i=1}
\left[
\log \phi\!\left(Z_{i,T};\mathbf W_i^\top\boldsymbol\beta+\delta_0+\delta_X X_i,\sigma_\epsilon^2\right)
+
\log(\pi)
+
\log \Phi(\mathbf{W}_i^\top\boldsymbol\gamma)
\right] \\
+&\sum_{i:D_i=0,\ Q_i=0}
\left[
\log(1-\pi)
+
\log \Phi(\mathbf{W}_i^\top\boldsymbol\gamma)
\right].
\end{align*}
}

As in the main-manuscript model, the regression component, the probit discontinuation component, and the retrieval component separate. In particular, estimation of $(\boldsymbol\beta,\delta_0,\delta_X,\sigma_\epsilon^2)$ reduces to a linear regression among subjects with observed final-visit outcomes using the additional interaction term $X_i \cdot \mathbf{1}(D_i=0)$.

\section{Derivation of observed log-likelihood}\label{subsec:obsl}

Since the endpoint measurements can be categorized into three groups--completers, RDs, and LTFUs--based on $\{(D_i,Q_i)\}_{i=1}^{n}$, the observed likelihood can be written as
\begin{equation*}
    \begin{aligned}
        \mathcal{L}_{\text{obs}} \left(\bfbeta, \delta,  \bfgamma, \pi, \sigma_{\epsilon}^2 \right)
        &= \prod_{i:D_{i}=1} P \left( Z_{i, T}, D_{i}=1 | \bfW_{i} ; \bfbeta, \sigma_{\epsilon}^2, \bfgamma \right)  \times \\
        & \prod_{i:(D_{i},Q_{i})=(0,1)} P \left( Z_{i, T}, D_{i}=0, Q_{i}=1 | \bfW_{i} ; \bfbeta, \delta, \sigma_{\epsilon}^2, \bfgamma, \pi \right)  \times \\
        & \prod_{i:(D_{i},Q_{i})=(0,0)} P \left( D_{i}=0, Q_{i}=0 | \bfW_{i} ; \bfgamma, \pi \right).
    \end{aligned}    
\end{equation*}
Note that the marginal likelihood of the observed data alone is captured by integrating out the unobserved or missing data, that is, 
\begin{equation*}
    \int P \left( Z_{i, T}, D_{i}=0,Q_{i}=0 | \bfW_{i} ; \bfbeta, \delta, \sigma_{\epsilon}^2, \bfgamma, \pi \right) \mathrm{d}Z_{i, T} = P \left( D_{i}=0,Q_{i}=0 | \bfW_{i} ; \bfgamma, \pi \right).
\end{equation*}

The probabilities in the observed likelihood can be explicitly expressed using the standard normal probability density and cumulative distribution functions, with different formulas derived based on the subject's status at the endpoint. For $D_{i}=1$, we have 
\begin{equation}\label{obsl}
    \begin{aligned}
    P \left( Z_{i, T}, D_{i}=1 | \bfW_{i} ; \bfbeta, \sigma_{\epsilon}^2 \right) &= P \left( Z_{i,T} | D_{i} = 1, \bfW_{i} ;  \bfbeta, \sigma_{\epsilon}^2 \right) P \left( D_{i} = 1 | \bfW_{i} ; \bfgamma \right) \\
    &= \phi\left(Z_{i,T} ; \bfW_{i}^\top\bfbeta,\sigma_{\epsilon}^{2}\right) \Phi\left(-\bfW_{i}^\top\bfgamma \right),
    \end{aligned}
\end{equation}
where $\phi(\cdot ; \mu,\sigma^{2})$ and $\Phi(\cdot ; \mu,\sigma^{2})$ denote the density and cumulative distribution functions of $N(\mu, \sigma^{2})$, respectively. For $(D_{i},Q_{i})=(0,1)$, it follows
\begin{equation*}
    \begin{aligned}
    P \big( Z_{i, T}, &D_{i}=0, Q_{i}=1 | \bfW_{i} ; \bfbeta, \sigma_{\epsilon}^2 \big)  \\
    = &P \left( Z_{i,T} | D_{i}=0, Q_{i}=1, \bfW_{i} ;  \bfbeta, \sigma_{\epsilon}^2 \right) P \left( Q_{i} = 1 | D_{i} = 0, \bfW_{i} ; \bfgamma \right) P \left( D_{i} = 0 | \bfW_{i} ; \bfgamma \right) \\
    = &\phi \left( Z_{i,T} ; \bfW_{i}^\top\bfbeta + \delta, \sigma_{\epsilon}^{2} \right) \cdot \pi \cdot \Phi \left( \bfW_{i}^\top\bfgamma \right).
    \end{aligned}
\end{equation*}
When $(D_{i},Q_{i})=(0,0)$, we have
\begin{equation*}
    \begin{aligned}
        P \left( D_{i}=0, Q_{i}=0 | \bfW_{i} ; \bfgamma, \pi \right)
        &=  P \left( Q_{i} = 0 | D_{i} = 0, \bfW_{i} ; \bfgamma \right) P \left( D_{i} = 0 | \bfW_{i} ; \bfgamma \right) \\
        &= (1-\pi) \cdot \Phi \left( \bfW_{i}^\top\bfgamma \right).
    \end{aligned}
\end{equation*}

Plugging the above into \eqref{obsl} gives the observed log-likelihood as
\begin{equation*}
    \begin{aligned}
        \ell_{\text{obs}} \left(\bfbeta, \delta,  \bfgamma, \pi, \sigma_{\epsilon}^2 \right)
        &= \sum_{i:D_{i}=1} \left\{ \log\phi\left(Z_{i,T} ; \bfW_{i}^\top\bfbeta,\sigma_{\epsilon}^{2}\right) + \log\Phi\left(-\bfW_{i}^\top\bfgamma \right) \right\} +\\
        & \sum_{i:(D_{i},Q_{i})=(0,1)} \left\{ \log\phi \left( Z_{i,T} ; \bfW_{i}^\top\bfbeta + \delta, \sigma_{\epsilon}^{2} \right) + \log(\pi) + \log\Phi \left( \bfW_{i}^\top\bfgamma \right) \right\} +\\
        & \sum_{i:(D_{i},Q_{i})=(0,0)} \left\{ \log(1-\pi) + \log\Phi \left( \bfW_{i}^\top\bfgamma \right) \right\}.
    \end{aligned}
\end{equation*}
By rearranging the terms, it becomes evident that $\ell_{\text{obs}} \left(\bfbeta, \delta,  \bfgamma, \pi, \sigma_{\epsilon}^2 \right)$ is the summation of three components: The first component is 
\begin{equation*}
    \sum_{i:D_{i}=1} \log\phi\left(Z_{i,T} ; \bfW_{i}^\top\bfbeta,\sigma_{\epsilon}^{2}\right) + \sum_{i:(D_{i},Q_{i})=(0,1)} \log\phi \left(Z_{i,T} ; \bfW_{i}^\top\bfbeta + \delta, \sigma_{\epsilon}^{2} \right)
\end{equation*}
which is independent of $\bfgamma$ and $\pi$. The second component is 
\begin{equation*}
    \sum_{i:D_{i}=1} \log\Phi\left(-\bfW_{i}^\top\bfgamma \right) + \sum_{i:D_{i}=0} \log\Phi \left( \bfW_{i}^\top\bfgamma \right)
\end{equation*}
which is independent of $\bfbeta$, $\delta$, $\sigma_{\epsilon}^2$ $\pi$. The third component is
\begin{equation*}
    \sum_{i:(D_{i},Q_{i})=(0,1)} \log(\pi) + \sum_{i:(D_{i},Q_{i})=(0,0)} \log(1-\pi)
\end{equation*}
which is independent of $\bfbeta$, $\delta$, $\sigma_{\epsilon}^2$ and $\bfgamma$.

\section{Derivation of ML estimators}\label{subsec:mle}

First, the ML estimators $\hat{\bfbeta}$ and $\hat{\delta}$ of $\bfbeta$ and $\delta$, respectively, can be derived by solving a least-squares optimization problem for standard linear regression, which is
\begin{equation*}
    \left[ \hat{\bfbeta}^\top , \hat{\delta} \right]^\top = \hat{\bfxi} = \underset{\bfxi}{\text{argmin}}\  \sum_{i:D_{i} = 1\ \text{or}\ (D_{i},Q_{i})=(0,1)} (Z_{i,T}-\bfU_{i}^\top \bfxi)^{2},
\end{equation*}
where $\bfU_{i} = \left[ \bfW_{i}^\top\ \mathbf{1}\{(D_{i},Q_{i})=(0,1)\} \right]^\top$ and $\bfxi = \left[ \bfbeta^\top , \delta \right]^\top$.     
The ML estimator of $\sigma_{\epsilon}^2$ is given by 
\[
\frac{1}{(N - \#[(D_{i},Q_{i})=(0,0)])} \sum_{i:D_{i} = 1\ \text{or}\ (D_{i},Q_{i})=(0,1)} (Z_{i,T}-\bfU_{i}^\top \hat{\bfxi})^{2},
\]
where $\#[(D_{i},Q_{i})=(0,0)]$ represents the number of subjects with $(D_{i},Q_{i})=(0,0)$. However, the denominator can be adjusted, as in standard linear regression, to obtain an unbiased estimator, that is,
\begin{equation*}
    \hat{\sigma}_{\epsilon}^2 = \frac{1}{(N-\#[(D_{i},Q_{i})=(0,0)]-\text{rank}(\bfU))} \sum_{i:D_{i} = 1\ \text{or}\ (D_{i},Q_{i})=(0,1)} (Z_{i,T}-\bfU_{i}^\top \hat{\bfxi})^{2},
\end{equation*}
where $\bfU$ is the design matrix including rows $\bfU_i$ for subjects with observed final-visit outcomes. If $\bfU$ is of full rank, then $\text{rank}(\bfU) = \text{ncol}(X)+3$. Denote by $\hat{\bfxi}_{j}$ the $j$-th element of $\hat{\bfxi}$. Based on $\hat{\sigma}_{\epsilon}$, the standard error of $\hat{\bfxi}_{j}$ can be computed as $\hat{\sigma}_{\epsilon} \sqrt{S_{jj}}$ where $S_{jj} = (\bfU^{\top} \bfU)^{-1}_{jj}$ is the $j$-th diagonal element of $(\bfU^\top\bfU)^{-1}$.
In practice, one can fit a linear regression model and obtain the resulting estimates and standard error by running the \texttt{lm()} function in \texttt{R}.

For $\bfgamma$, the ML estimator $\hat{\bfgamma}$ can be obtained by solving 
\begin{equation*}
    \hat{\bfgamma} = \underset{\bfgamma}{\text{argmax}} \sum_{i=1}^{N} \left\{D_i \log\Phi(-\bfW_{i}^\top\bfgamma) + (1-D_i)\log\Phi(\bfW_{i}^\top\bfgamma)\right\}.
\end{equation*} 
This is an optimization problem for fitting the classical probit model, which satisfies the regularity conditions required for the maximum likelihood estimator $\bfgamma$ to be consistent and asymptotically normally distributed, See Section 15.3 of \cite{Davidson1993}. In practice, one can fit a probit model using the \texttt{glm()} function with \texttt{family = binomial(link = "probit")} in \texttt{R}.
For $\pi$, the ML estimate can be readily obtained by calculating the proportion of subjects with $(D_{i},Q_{i}) = (0,1)$ relative to those with $D_{i} = 0$ as follows:
\begin{equation*}
    \hat{\pi} = \frac{\#[(D_{i},Q_{i}) = (0,1)]}{ \#[D_{i} = 0]}.
\end{equation*}

\section{Implementation of existing methods}\label{subsec:cmpt}

We considered three existing imputation methods commonly used for the estimand under the TP strategy: return-to-baseline (RTB) imputation \citep{Qu2022}, washout imputation, and RD imputation \citep{Wang2022}, with implementations based on the formulations described in literature \cite{Wang2023}.

RTB imputation assumes that the treatment effect is fully lost immediately upon treatment discontinuation, and the subject's response returns to their baseline level. This method is more commonly applied in active-controlled trials. In our implementation, each missing value was imputed by sampling from a normal distribution with mean equal to the corresponding subject's baseline value and standard deviation estimated from an ANCOVA model of the endpoint, using baseline measurement and treatment group as covariates. RD data were not treated differently from completers and were included directly in the model.

Washout imputation is based on the conservative assumption that the treatment effect does not persist after treatment discontinuation. This assumption is often used in primary or sensitivity analyses, particularly for the estimand under TP strategy or in regulatory settings where a worst-case scenario is preferred. The washout approach is generally considered appropriate for placebo-controlled trials. From a statistical perspective, it treats RD data as contaminated due to noncompliance (i.e., treatment discontinuation) and therefore considers them missing. In our implementation, missing values were imputed by sampling from a normal distribution with mean equal to the average endpoint value among placebo subjects and standard deviation estimated from an ANCOVA model fitted to endpoint data from placebo completers, using baseline measurement as a covariate.

RD imputation, in contrast, assumes that endpoint assessments observed in RDs are representative of those that would have been observed in unretrieved dropouts, conditional on observed covariates--an assumption closely aligned with that underlying our proposed method. In our implementation, missing endpoint values were imputed by sampling from a normal distribution with mean equal to the average endpoint value among RDs and standard deviation estimated from an ANCOVA model using RD data only, with baseline measurement and treatment group as covariates. Like our method, the RD approach implicitly assumes that RDs behave similarly to non-responders after treatment discontinuation.

\section{Additional simulation results}\label{subsec:simout}

Tables \ref{tab:simout3} and \ref{tab:simout4} present simulation results for the smaller sample size $N = 100$ under the same treatment-effect configurations considered in the main manuscript, with retrieval probability $\pi = 0.5$. As expected, all methods are less precise than in the corresponding $N = 200$ settings reported in Tables \ref{tab:simout1} and \ref{tab:simout2} of the main manuscript. Nevertheless, the qualitative conclusions remain unchanged. In the efficacious scenarios, RTB and washout imputation continue to show substantial positive bias, whereas RD imputation and the proposed method remain essentially unbiased. RTB may attain a smaller RMSE than the proposed method in some settings, but this occurs despite appreciable bias and does not translate into better inferential calibration. Relative to RD imputation, the proposed method generally reduces RMSE, yields higher rejection rates, and provides empirical coverage closer to the nominal 95\% level, with CI lengths that are competitive.

When the placebo arm has the higher treatment-discontinuation rate in Table \ref{tab:simout3}, the proposed method performs particularly well in terms of inferential calibration. In Scenarios 1 and 2, it combines near-zero bias with lower RMSE than RD imputation, higher rejection rates, and coverage closer to 95\%. In Scenario 3, all methods have small bias, but the proposed method again yields rejection rate and coverage closest to the nominal levels. When the experimental arm has the higher treatment-discontinuation rate (Table \ref{tab:simout4}), the same general pattern is observed. RD imputation remains nearly unbiased but exhibits larger RMSE and under-coverage, whereas the proposed method improves coverages and rejection rates while maintaining similarly small bias. These results show that the conclusions from the main manuscript persist under a smaller sample size.

Tables \ref{tab:simout5} and \ref{tab:simout6} examine a lower retrieval probability, $\pi = 0.3$, for the two discontinuation configurations. The same qualitative pattern persists, but the gap between RD imputation and the proposed method becomes more pronounced. Across the six scenarios in Tables \ref{tab:simout5} and \ref{tab:simout6}, RD imputation shows noticeable RMSE inflation and empirical coverage ranging from 87.84\% to 89.98\%, together with inflated empirical rejection rate in Scenario 3. By contrast, the proposed method remains essentially unbiased, attains the highest rejection rates in the efficacious scenarios, and keeps empirical coverages much closer to the nominal level, ranging from 94.36\% to 95.42\%. Although its CIs become somewhat longer than those of RTB or washout imputation in some settings, this appears to be the cost of preserving inferential calibration as the number of observed RDs declines.

Tables \ref{tab:simout7} and \ref{tab:simout8} consider the more extreme case $\pi = 0.1$. Here the deterioration of RD imputation becomes substantial. Even though average bias remains relatively small, RD imputation exhibits severe RMSE inflation, with RMSE ranging from 6.697 to 7.675 in Table \ref{tab:simout7} and from 7.380 to 9.443 in Table \ref{tab:simout8}. Its empirical coverage drops sharply to 76.33\%-77.74\% when treatment discontinuation is more frequent in the placebo arm and to 70.50\%-72.31\% when treatment discontinuation is more frequent in the experimental arm; in Scenario 3, the empirical type I error rises to 0.223 and 0.284, respectively. By contrast, the proposed method remains close to unbiased across scenarios, achieves the highest rejection rates in the efficacious scenarios, and maintains empirical coverage between 94.68\% and 95.32\%, with type I error near the nominal 0.05 level. Although the proposed confidence intervals are longer than those of RTB or washout imputation when $\pi = 0.1$, this increase is modest relative to the pronounced instability of RD imputation and is accompanied by markedly better inferential calibration.

Taken together, Tables \ref{tab:simout3}-\ref{tab:simout8} show that the proposed method is substantially less sensitive than RD imputation to the availability of observed RD data. The main vulnerability of RD imputation in these simulations is not systematic point-estimation bias, but the instability induced by relying on a limited amount of RD data. Since the proposed method borrows information through a joint model for the endpoint and discontinuation processes, it retains stable estimation and near-nominal inference even when the retrieval probability is low.

\newpage
\renewcommand{\arraystretch}{1.1}
\begin{table}[t]
\centering
\caption{Performance of each method under three scenarios with higher treatment discontinuation rate in the placebo arm ($\gamma_X = -0.25$) and sample size $N=100$}
\vspace{1em}
\begin{tabular}{@{}lcccccc@{}}
\toprule \toprule
\multicolumn{1}{c}{\multirow{2}{*}{Method}} &
  \multirow{2}{*}{$\beta_X^{\text{TP}}$} &
  \multirow{2}{*}{Bias} &
  \multirow{2}{*}{RMSE} &
  \multirow{2}{*}{Rejection Rate} &
  \multicolumn{2}{c}{95$\%$ CI} \\ \cmidrule(l){6-7}
\multicolumn{1}{c}{} & & & & & Coverage ($\%$) & Length \\ \midrule

\multicolumn{7}{l}{Scenario 1: Efficacious drug with persisting effect ($\beta_X = -10$ and $\delta = 5$)} \\ \midrule
RTB imputation       & \multirow{4}{*}{-10.291} & \phantom{ }1.330 & 4.007 & 0.556 & 96.16 & 16.862 \\
Washout imputation   &                          & \phantom{ }2.679 & 4.530 & 0.363 & 95.46 & 17.809 \\
RD imputation        &                          & \phantom{ }0.009 & 4.721 & 0.648 & 92.40 & 16.997 \\
Our method           &                          & -0.008 & 4.362 & 0.669 & 94.20 & 16.789 \\ \midrule

\multicolumn{7}{l}{Scenario 2: Efficacious drug without persisting effect ($\beta_X = -10$ and $\delta = 10$)} \\ \midrule
RTB imputation       & \multirow{4}{*}{-10.582} & \phantom{ }1.416 & 4.020 & 0.567 & 96.58 & 17.000 \\
Washout imputation   &                          & \phantom{ }2.911 & 4.673 & 0.369 & 94.22 & 17.797 \\
RD imputation        &                          & -0.137 & 4.729 & 0.671 & 93.02 & 17.161 \\
Our method           &                          & -0.080 & 4.355 & 0.682 & 94.52 & 16.962 \\ \midrule

\multicolumn{7}{l}{Scenario 3: Inefficacious drug ($\beta_X = 0$ and $\delta = 0$)} \\ \midrule
RTB imputation       & \multirow{4}{*}{0.000}   & \phantom{ }0.008 & 3.695 & 0.025 & 97.50 & 16.741 \\
Washout imputation   &                         & \phantom{ }0.017 & 3.587 & 0.016 & 98.44 & 17.680 \\
RD imputation        &                         & -0.016 & 4.604 & 0.073 & 92.72 & 16.934 \\
Our method           &                         & \phantom{ }0.014 & 4.294 & 0.056 & 94.12 & 16.741 \\ \bottomrule \bottomrule
\end{tabular}
\label{tab:simout3}
\end{table}

\renewcommand{\arraystretch}{1.1}
\renewcommand{\arraystretch}{1.1}
\begin{table}[t]
\centering
\caption{Performance of each method under three scenarios with higher treatment discontinuation rate in the experimental arm ($\gamma_X = 0.25$) and sample size $N=100$}
\vspace{1em}
\begin{tabular}{@{}lcccccc@{}}
\toprule \toprule
\multicolumn{1}{c}{\multirow{2}{*}{Method}} &
  \multirow{2}{*}{$\beta_X^{\text{TP}}$} &
  \multirow{2}{*}{Bias} &
  \multirow{2}{*}{RMSE} &
  \multirow{2}{*}{Rejection Rate} &
  \multicolumn{2}{c}{95$\%$ CI} \\ \cmidrule(l){6-7}
\multicolumn{1}{c}{} & & & & & Coverage ($\%$) & Length \\ \midrule

\multicolumn{7}{l}{Scenario 1: Efficacious drug with persisting effect ($\beta_X = -10$ and $\delta = 5$)} \\ \midrule
RTB imputation       & \multirow{4}{*}{-9.681} & \phantom{ }1.682 & 4.082 & 0.436 & 96.54 & 17.097 \\
Washout imputation   &                          & \phantom{ }3.299 & 4.593 & 0.198 & 96.24 & 18.248 \\
RD imputation        &                          & \phantom{ }0.069 & 4.842 & 0.586 & 92.10 & 17.143 \\
Our method           &                          & \phantom{ }0.050 & 4.442 & 0.588 & 94.30 & 17.064 \\ \midrule

\multicolumn{7}{l}{Scenario 2: Efficacious drug without persisting effect ($\beta_X = -10$ and $\delta = 10$)} \\ \midrule
RTB imputation       & \multirow{4}{*}{-9.361} & \phantom{ }1.478 & 4.072 & 0.418 & 96.68 & 17.275 \\
Washout imputation   &                          & \phantom{ }2.949 & 4.377 & 0.201 & 97.06 & 18.243 \\
RD imputation        &                          & \phantom{ }0.020 & 4.915 & 0.547 & 92.08 & 17.346 \\
Our method           &                          & -0.008 & 4.543 & 0.551 & 94.34 & 17.270 \\ \midrule

\multicolumn{7}{l}{Scenario 3: Inefficacious drug ($\beta_X = 0$ and $\delta = 0$)} \\ \midrule
RTB imputation       & \multirow{4}{*}{0.00}   & \phantom{ }0.025 & 3.644 & 0.025 & 97.54 & 16.974 \\
Washout imputation   &                         & \phantom{ }0.004 & 3.160 & 0.008 & 99.22 & 18.075 \\
RD imputation        &                         & \phantom{ }0.064 & 4.744 & 0.073 & 92.66 & 17.110 \\
Our method           &                         & \phantom{ }0.041 & 4.383 & 0.056 & 94.14 & 17.030 \\ \bottomrule \bottomrule
\end{tabular}
\label{tab:simout4}
\end{table}

\renewcommand{\arraystretch}{1.1}
\begin{table}[t]
\centering
\caption{Performance of each method under three scenarios with $\pi = 0.3$, higher treatment discontinuation rate in the placebo arm ($\gamma_X = -0.25$) and sample size $N=200$}
\vspace{1em}
\begin{tabular}{@{}lcccccc@{}}
\toprule \toprule
\multicolumn{1}{c}{\multirow{2}{*}{Method}} &
  \multirow{2}{*}{$\beta_X^{\text{TP}}$} &
  \multirow{2}{*}{Bias} &
  \multirow{2}{*}{RMSE} &
  \multirow{2}{*}{Rejection Rate} &
  \multicolumn{2}{c}{95$\%$ CI} \\ \cmidrule(l){6-7}
\multicolumn{1}{c}{} & & & & & Coverage ($\%$) & Length \\ \midrule

\multicolumn{7}{l}{Scenario 1: Efficacious drug with persisting effect ($\beta_X = -10$ and $\delta = 5$)} \\ \midrule
RTB imputation       & \multirow{4}{*}{-10.291} & \phantom{ }1.861 & 3.189 & 0.815 & 94.72 & 12.198 \\
Washout imputation   &                          & \phantom{ }2.679 & 3.711 & 0.689 & 91.70 & 12.595 \\
RD imputation        &                          & -0.002 & 3.833 & 0.858 & 88.94 & 12.261 \\
Our method           &                          & -0.023 & 3.188 & 0.902 & 94.36 & 12.344 \\ \midrule

\multicolumn{7}{l}{Scenario 2: Efficacious drug without persisting effect ($\beta_X = -10$ and $\delta = 10$)} \\ \midrule
RTB imputation       & \multirow{4}{*}{-10.582} & \phantom{ }2.069 & 3.312 &  0.826 & 94.08 & 12.264 \\
Washout imputation   &                          & \phantom{ }2.980 & 3.922 & 0.700 & 90.10 & 12.580 \\
RD imputation        &                          & -0.072 & 3.889 & 0.882 & 89.02 & 12.389 \\
Our method           &                          & -0.030 & 3.194 & 0.917 & 94.70 & 12.463 \\ \midrule

\multicolumn{7}{l}{Scenario 3: Inefficacious drug ($\beta_X = 0$ and $\delta = 0$)} \\ \midrule
RTB imputation       & \multirow{4}{*}{0.000}   & -0.019 & 2.483 &  0.013 & 98.66 & 12.091 \\
Washout imputation   &                         & -0.020 & 2.479 &  0.011 & 98.86 & 12.490 \\
RD imputation        &                         & -0.027 & 3.745 & 0.100 & 89.98 & 12.230 \\
Our method           &                         & -0.014 & 3.085 & 0.044 & 95.42 & 12.289 \\ \bottomrule \bottomrule
\end{tabular}
\label{tab:simout5}
\end{table}

\renewcommand{\arraystretch}{1.1}
\begin{table}[t]
\centering
\caption{Performance of each method under three scenarios with $\pi = 0.3$, higher treatment discontinuation rate in the experimental arm ($\gamma_X = 0.25$) and sample size $N=200$}
\vspace{1em}
\begin{tabular}{@{}lcccccc@{}}
\toprule \toprule
\multicolumn{1}{c}{\multirow{2}{*}{Method}} &
  \multirow{2}{*}{$\beta_X^{\text{TP}}$} &
  \multirow{2}{*}{Bias} &
  \multirow{2}{*}{RMSE} &
  \multirow{2}{*}{Rejection Rate} &
  \multicolumn{2}{c}{95$\%$ CI} \\ \cmidrule(l){6-7}
\multicolumn{1}{c}{} & & & & & Coverage ($\%$) & Length \\ \midrule

\multicolumn{7}{l}{Scenario 1: Efficacious drug with persisting effect ($\beta_X = -10$ and $\delta = 5$)} \\ \midrule
RTB imputation       & \multirow{4}{*}{-9.681} & \phantom{ }2.291 & 3.409 & 0.674 & 93.54 & 12.421 \\
Washout imputation   &                          & \phantom{ }3.280 & 3.988 & 0.480 & 91.12 & 12.888 \\
RD imputation        &                          & -0.040 & 3.965 & 0.810 & 88.36 & 12.464 \\
Our method           &                          & -0.025 & 3.262 & 0.840 & 94.62 & 12.640 \\ \midrule

\multicolumn{7}{l}{Scenario 2: Efficacious drug without persisting effect ($\beta_X = -10$ and $\delta = 10$)} \\ \midrule
RTB imputation       & \multirow{4}{*}{-9.361} & \phantom{ }2.096 & 3.270 & 0.652 & 94.98 & 12.541 \\
Washout imputation   &                          & \phantom{ }2.985 & 3.739 & 0.483 & 93.62 & 12.921 \\
RD imputation        &                          & \phantom{ }0.045 & 3.997 & 0.777 & 88.36 & 12.624 \\
Our method           &                          & \phantom{ }0.010 & 3.282 & 0.811 & 94.62 & 12.805 \\ \midrule

\multicolumn{7}{l}{Scenario 3: Inefficacious drug ($\beta_X = 0$ and $\delta = 0$)} \\ \midrule
RTB imputation       & \multirow{4}{*}{0.000}   & -0.043 & 2.479 &  0.014 & 98.64 & 12.306 \\
Washout imputation   &                         & -0.037 & 2.216 &  0.003 & 99.70 & 12.780 \\
RD imputation        &                         & -0.036 & 3.964 & 0.122 & 87.84 & 12.421 \\
Our method           &                         & -0.062 & 3.251 & 0.055 & 94.38 & 12.606 \\ \bottomrule \bottomrule
\end{tabular}
\label{tab:simout6}
\end{table}

\renewcommand{\arraystretch}{1.1}
\begin{table}[t]
\centering
\caption{Performance of each method under three scenarios with $\pi = 0.1$, higher treatment discontinuation rate in the placebo arm ($\gamma_X = -0.25$) and sample size $N=200$}
\vspace{1em}
\begin{tabular}{@{}lcccccc@{}}
\toprule \toprule
\multicolumn{1}{c}{\multirow{2}{*}{Method}} &
  \multirow{2}{*}{$\beta_X^{\text{TP}}$} &
  \multirow{2}{*}{Bias} &
  \multirow{2}{*}{RMSE} &
  \multirow{2}{*}{Rejection Rate} &
  \multicolumn{2}{c}{95$\%$ CI} \\ \cmidrule(l){6-7}
\multicolumn{1}{c}{} & & & & & Coverage ($\%$) & Length \\ \midrule

\multicolumn{7}{l}{Scenario 1: Efficacious drug with persisting effect ($\beta_X = -10$ and $\delta = 5$)} \\ \midrule
RTB imputation       & \multirow{4}{*}{-10.291} & \phantom{ }2.395 & 3.467 & 0.746 & 93.48 & 12.473 \\
Washout imputation   &                          & \phantom{ }2.677 & 3.709 & 0.691 & 91.82 & 12.595 \\
RD imputation        &                          & \phantom{ }0.299 & 6.697 & 0.757 & 76.33 & 13.327 \\
Our method           &                          & -0.031 & 3.361 & 0.861 & 94.71 & 13.195 \\ \midrule

\multicolumn{7}{l}{Scenario 2: Efficacious drug without persisting effect ($\beta_X = -10$ and $\delta = 10$)} \\ \midrule
RTB imputation       & \multirow{4}{*}{-10.582} & \phantom{ }2.670 & 3.646 & 0.752 & 92.19 & 12.489 \\
Washout imputation   &                          & \phantom{ }2.978 & 3.922 & 0.698 & 90.10 & 12.582 \\
RD imputation        &                          & \phantom{ }0.193 & 7.675 & 0.777 & 77.29 & 13.415 \\
Our method           &                          & -0.023 & 3.342 & 0.880 & 95.18 & 13.305 \\ \midrule

\multicolumn{7}{l}{Scenario 3: Inefficacious drug ($\beta_X = 0$ and $\delta = 0$)} \\ \midrule
RTB imputation       & \multirow{4}{*}{0.000}   & -0.023 & 2.408 &  0.011 & 98.85 & 12.363 \\
Washout imputation   &                         & -0.021 & 2.480 &  0.012 & 98.81 & 12.491 \\
RD imputation        &                         & \phantom{ }0.032 & 6.789 & 0.223 & 77.74 & 13.325 \\
Our method           &                         & -0.030 & 3.281 & 0.047 & 95.32 & 13.145 \\ \bottomrule \bottomrule
\end{tabular}
\label{tab:simout7}
\end{table}

\renewcommand{\arraystretch}{1.1}
\begin{table}[t]
\centering
\caption{Performance of each method under three scenarios with $\pi = 0.1$, higher treatment discontinuation rate in the experimental arm ($\gamma_X = 0.25$) and sample size $N=200$}
\vspace{1em}
\begin{tabular}{@{}lcccccc@{}}
\toprule \toprule
\multicolumn{1}{c}{\multirow{2}{*}{Method}} &
  \multirow{2}{*}{$\beta_X^{\text{TP}}$} &
  \multirow{2}{*}{Bias} &
  \multirow{2}{*}{RMSE} &
  \multirow{2}{*}{Rejection Rate} &
  \multicolumn{2}{c}{95$\%$ CI} \\ \cmidrule(l){6-7}
\multicolumn{1}{c}{} & & & & & Coverage ($\%$) & Length \\ \midrule

\multicolumn{7}{l}{Scenario 1: Efficacious drug with persisting effect ($\beta_X = -10$ and $\delta = 5$)} \\ \midrule
RTB imputation       & \multirow{4}{*}{-9.681} & \phantom{ }2.937 & 3.811 & 0.559 & 91.89 & 12.741 \\
Washout imputation   &                          & \phantom{ }3.279 & 3.986 & 0.482 & 91.03 & 12.889 \\
RD imputation        &                          & -0.105 & 9.443 & 0.706 & 70.50 & 13.629 \\
Our method           &                          & -0.040 & 3.480 & 0.792 & 94.68 & 13.570 \\ \midrule

\multicolumn{7}{l}{Scenario 2: Efficacious drug without persisting effect ($\beta_X = -10$ and $\delta = 10$)} \\ \midrule
RTB imputation       & \multirow{4}{*}{-9.361} & \phantom{ }2.694 & 3.613 & 0.542 & 93.48 & 12.815 \\
Washout imputation   &                          & \phantom{ }2.987 & 3.740 & 0.483 & 93.78 & 12.919 \\
RD imputation        &                          & \phantom{ }0.248 & 7.554 & 0.683 & 72.31 & 13.748 \\
Our method           &                          & \phantom{ }0.016 & 3.479 & 0.759 & 95.12 & 13.729 \\ \midrule

\multicolumn{7}{l}{Scenario 3: Inefficacious drug ($\beta_X = 0$ and $\delta = 0$)} \\ \midrule
RTB imputation       & \multirow{4}{*}{0.000}   & -0.037 & 2.365 &  0.007 & 99.34 & 12.626 \\
Washout imputation   &                         & -0.040 & 2.216 &  0.003 & 99.70 & 12.782 \\
RD imputation        &                         & \phantom{ }0.123 & 7.380 & 0.284 & 71.62 & 13.482 \\
Our method           &                         & -0.068 & 3.431 & 0.049 & 94.92 & 13.533 \\ \bottomrule \bottomrule
\end{tabular}
\label{tab:simout8}
\end{table}

\clearpage
\section{Additional results for Section 4}\label{subsec:appout_add}

\begin{figure}[t]
    \centering
    \includegraphics[width=1\textwidth]{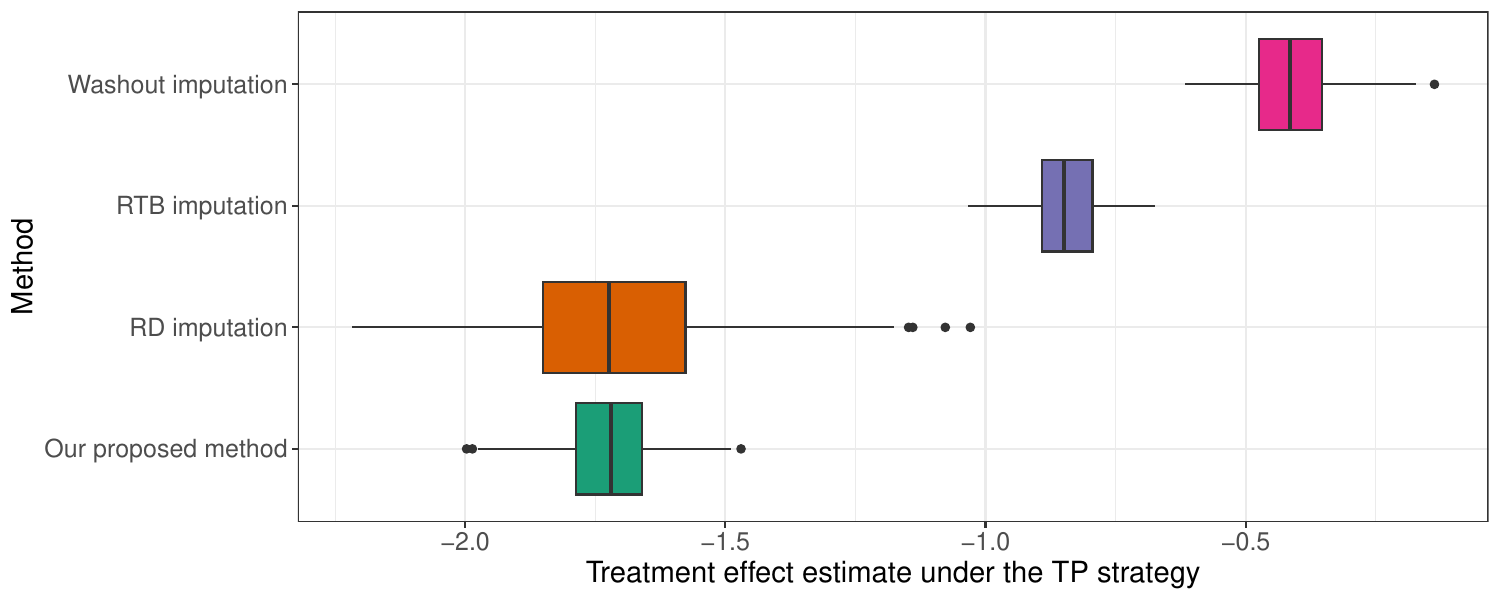}
    \caption{Treatment effect estimates under the TP strategy based on 500 synthetic HAMD17 datasets generated from the original trial data under no treatment effect persistence after discontinuation}
    \label{fig:application_add}
\end{figure}

Figure \ref{fig:application_add} presents numerical results for the same illustrative application considered in the main manuscript, but under the assumption of no persisting treatment effect after treatment discontinuation. Specifically, in the replacement step of the structured synthetic RD generation procedure based on the two-step approach \cite{Wang2023}, a multiplier of 0 is used. As noted in the main manuscript, the proposed method is close to RD imputation in the location of the estimates, although RD imputation still exhibits greater variability than the proposed method and the other competing methods. Taken together, Figures \ref{fig:application} and \ref{fig:application_add} suggest that RD imputation can be sensitive when the observed RDs are not representative of unretrieved discontinuers, as in the structured tail-based RD generation scheme considered here, or when the amount of observed RD data is limited. By contrast, the proposed method appears less sensitive to these features because it uses RD information through a joint model for the endpoint and discontinuation processes rather than through direct donor-based imputation.

\end{document}